\documentclass{article}

\usepackage[final]{aaj2026}
\usepackage[numbers]{natbib}
\usepackage{bm}
\usepackage{amssymb}
\usepackage{booktabs}
\usepackage{multirow}
\usepackage{graphicx}
\usepackage{siunitx}
\usepackage{amsmath}
\usepackage{bbm}
\usepackage{graphicx}

\newcommand{\topone}[1]{\bfseries #1}        
\newcommand{\toptwo}[1]{\underline{#1}}      

\newcommand{\m}[4]{#1 & #2 & #3 & #4}

\usepackage[utf8]{inputenc} 
\usepackage[T1]{fontenc}    
\usepackage{hyperref}       
\usepackage{url}            
\usepackage{booktabs}       
\usepackage{amsfonts}       
\usepackage{nicefrac}       
\usepackage{microtype}      
\usepackage{xcolor}         

\title{Continual Learning for fMRI-Based Brain Disorder Diagnosis via Functional Connectivity Matrices Generative Replay}

\author{Qianyu Chen$^{\dagger}$\\
$^{\dagger}$Nanyang Technological University\\
{\tt\small CHEN1981@e.ntu.edu.sg}
\And
Shujian Yu$^{\ddagger,\S}$\\
$^{\ddagger}$VU Amsterdam;
$^\S$UiT The Arctic University of Norway\\
{\tt\small yusj9011@gmail.com}
}

\begin{document}

\maketitle

\begin{abstract}
  Functional magnetic resonance imaging (fMRI) is widely used for studying and diagnosing brain disorders, with functional connectivity (FC) matrices providing powerful representations of large-scale neural interactions. However, existing diagnostic models are trained either on a single site or under full multi-site access, making them unsuitable for real-world scenarios where clinical data arrive sequentially from different institutions. This results in limited generalization and severe catastrophic forgetting.
  This paper presents the first continual learning framework specifically designed for fMRI-based diagnosis across heterogeneous clinical sites. Our framework introduces a structure-aware variational autoencoder that synthesizes realistic FC matrices for both patient and control groups. Built on this generative backbone, we develop a multi-level knowledge distillation strategy that aligns predictions and graph representations between new-site data and replayed samples. To further enhance efficiency, we incorporate a hierarchical contextual bandit scheme for adaptive replay sampling.
  Experiments on multi-site datasets for major depressive disorder (MDD), schizophrenia (SZ), and autism spectrum disorder (ASD) show that the proposed generative model enhances data augmentation quality, and the overall continual learning framework substantially outperforms existing methods in mitigating catastrophic forgetting. Our code is available at \url{https://github.com/4me808/FORGE}.
\end{abstract}

\section{Introduction}
\label{sec:intro}

Functional magnetic resonance imaging (fMRI) is a non-invasive technique that measures brain activity through blood-oxygen-level-dependent (BOLD) signals. Functional connectivity (FC) matrices derived from these signals encode pairwise correlations between brain regions, characterizing large-scale network organization and inter-regional interaction patterns. As such, FC provides a principled representation of brain network topology and a powerful modality for probing neurobiological mechanisms and disease-related alterations~\citep{plitt2015functional,ASD,gallo2023functional}. 
FC matrices, combined with graph-based modeling techniques~\citep{kawahara2017brainnetcnn,ASD}, have been widely employed for brain disorder diagnosis. However, existing fMRI-based diagnostic models still exhibit limited reliability for practical clinical deployment~\citep{plitt2015functional,desai2021continual,amrollahi2022leveraging}. Clinical data typically arrive sequentially from different sites, yet most models are trained offline on a single dataset or assume immediate access to all data. Consequently, they lack mechanisms for continual generalization to newly arriving data and suffer from the overwriting of previously learned information, a phenomenon known as \emph{catastrophic forgetting}~\citep{parisi2019continual,van2019three}. When such models are adapted to new sites, updates to accommodate new distributions often overwrite prior knowledge, substantially undermining their clinical utility in real-world deployment~\citep{mccloskey1989catastrophic,Kirkpatrick_2017,desai2021continual}. This limitation motivates the development of continual learning (CL) frameworks that can integrate new-site data incrementally while preserving previously acquired knowledge~\citep{parisi2019continual,amrollahi2022leveraging}.


Various strategies have been proposed to preserve knowledge from previous tasks in CL, with notable examples including Elastic Weight Consolidation (EWC)~\citep{Kirkpatrick_2017} and PackNet~\citep{mallya2018packnet}. 
Among them, rehearsal-based approaches alleviate forgetting by explicitly storing and replaying a few representative samples from previous tasks in a memory buffer.
In healthcare, however, experience replay is often constrained by privacy regulations that prohibit sharing raw patient data across institutions~\citep{chaudhry2019tiny,gkoulalas2015privacy,desai2021continual}. Furthermore, most CL research has focused on image-based tasks, leaving graph-structured medical data, especially fMRI, largely underexplored~\citep{parisi2019continual,van2019three,ASD,10680255,zeng2019multi}. These considerations highlight the urgent need for a graph-specific, privacy-friendly CL framework tailored for streaming fMRI data.

In this paper, we propose FORGE (Functional cOnnectivity Replay with Graph rEpresentation), a CL framework tailored for FC matrices collected across multiple clinical sites. FORGE addresses catastrophic forgetting by replaying both low-level FC matrices and higher-level knowledge in the form of classification logits and graph embeddings within a unified knowledge distillation framework. A central contribution of this work is FCM-VAE, a novel generative model specifically designed for FC matrices. FCM-VAE enables privacy-preserving generative replay by synthesizing realistic FC patterns without exposing raw patient data. Unlike existing VAEs, it employs a structure-aware graph transformer encoder with local adjacency encoding and spectral positional encoding to capture the intrinsic topological and spectral properties of FC networks. Moreover, we introduce a low-rank decoder, which is well suited for FC matrices because large-scale functional connectivity is known to exhibit strong low-rank structure~\citep{saggar2018towards}. 






Our main contributions include:
\begin{itemize}

\item We propose FORGE, the first CL framework tailored for cross-site fMRI-based disease diagnosis under realistic streaming and privacy-friendly settings. FORGE unifies dual-level knowledge distillation with generative replay.

\item We develop FCM-VAE, a conditional variational autoencoder specifically designed for FC matrices. FCM-VAE synthesizes realistic and diverse brain network patterns by integrating local adjacency encoding, spectral positional encoding, and a low-rank decoder, and it outperforms existing graph generative models in sample fidelity.

\item Extensive experiments on large-scale neuroimaging datasets, including ABIDE ~\citep{ASD}, REST-meta-MDD ~\citep{MDD}, and BSNIP ~\citep{BSNIP}, show that FCM-VAE significantly improves augmentation quality, and FORGE consistently exceeds state-of-the-art baselines in predictive accuracy and resistance to catastrophic forgetting.

\end{itemize}

\section{RELATED WORK}\label{related work}
\subsection{Continual Learning and its Extensions to Graph Data}

Existing continual learning (CL) approaches generally fall into four main categories. Regularization-based methods constrain parameter updates or enforce consistency with earlier predictions to preserve prior knowledge~\citep{Kirkpatrick_2017,li2017learning}. Parameter-isolation methods allocate task-specific model capacity to reduce interference and forgetting~\citep{rusu2016progressive,mallya2018packnet,yoon2018lifelong}. Rehearsal-based strategies replay real or synthetic samples from previous tasks to maintain performance over time~\citep{rebuffi2017icarl,chaudhry2019efficient,buzzega2020dark,kim2024sddgr}. Knowledge-distillation methods adopt a teacher–student paradigm, guiding the current model to align with the outputs or representations of earlier models~\citep{10204250,song2024llmbasedprivacydataaugmentation,zhang2025continualdistillationlearningknowledge}.
However, most CL frameworks are developed for image-based data. When applied to graph-structured domains, they often struggle with non-Euclidean topologies and evolving relational structures, limiting their ability to prevent forgetting and maintain generalization~\citep{Tian2024ContinualLO}. 


Recent work has begun to develop CL frameworks for graph data~\citep{zhang2022hpn,wang2023sgnngr}. These approaches typically extend conventional CL paradigms with graph-aware mechanisms such as structure-guided replay, topology-regularized updates, and subgraph sampling, aiming to improve node representations and retain knowledge over time~\citep{zhou2023ergnn,wang2023sgnngr,10.1145/3702648,Zhang_2025}. However, most existing studies focus on node-level classification, while graph-level classification remains largely underexplored~\citep{Tian2024ContinualLO}. A few works address graph continual classification using basic rehearsal strategies that replay stored graphs to alleviate forgetting~\citep{hoang2023universalgraphcontinuallearning,carta2021catastrophicforgettingdeepgraph}, or by preserving topology-aware features to reduce representation drift~\citep{Zhang_2024,liu2020overcomingcatastrophicforgettinggraph}. Notably, none of these methods have been extended to or evaluated on brain network classification tasks.




\subsection{Continual Learning for Brain Signals}
Several recent studies have begun exploring CL for neural recording signals, but the literature has largely focused on subject-specific adaptation for EEG. For example, \citet{SHI2024109028} proposes a memory projection strategy that aligns gradients between new and old tasks through dynamic projection matrices, reducing forgetting in seizure prediction. \citet{DUAN2024106338} introduces a balanced replay buffer that maintains a small class-balanced subset of samples to alleviate forgetting in online EEG decoding. More recently, \citet{zhou2025brainuicl} presents BrainUICL, an unsupervised framework that adapts to new subjects without labels while preserving performance on previously observed subjects.
In contrast, our work addresses a substantially more challenging cross-site setting for fMRI, where heterogeneity across institutions, scanners, and acquisition protocols introduces far greater distributional shift. Our goal is to mitigate forgetting under this site-level variability while ensuring accuracy, generalization, and privacy preservation in practical deployment. From a neuroscience perspective, fMRI and EEG differ markedly in their spatial resolution, temporal dynamics~\citep{logothetis2008we}, and network-level representations, implying fundamentally different challenges for CL.

\section{FORGE}\label{Forge}

\subsection{Problem Setup}
We first describe how resting-state fMRI (rs-fMRI) data are transformed into FC matrices for classification, and then formalize the diagnosis task under CL.
\begin{figure}[htbp]
    \centering
    \includegraphics[width=0.75\textwidth]{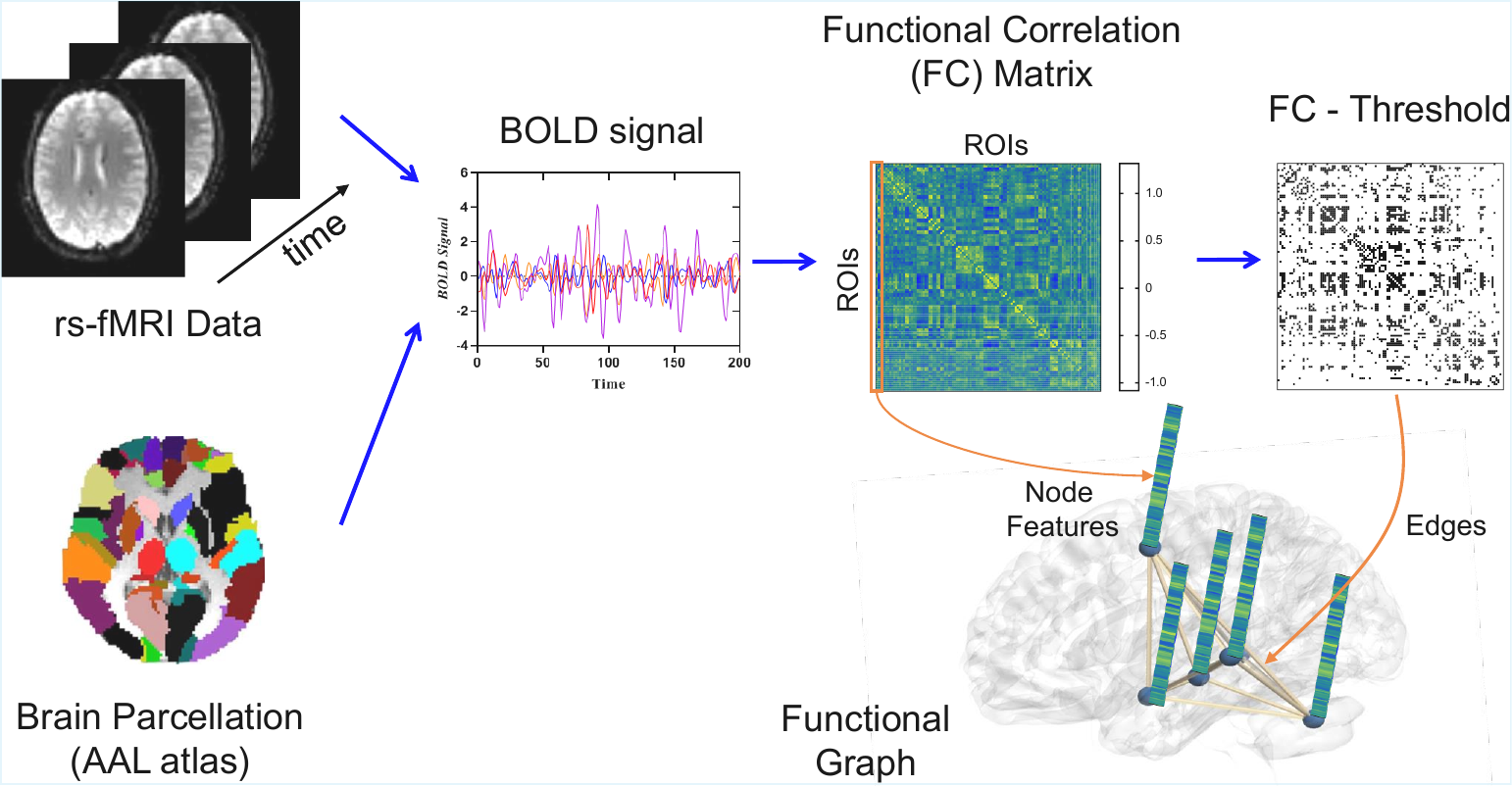}
    \caption{Pipeline for transforming raw rs-fMRI data into brain networks for the classification task.}
    \label{fig:fMRI}
\end{figure}
\noindent\textbf{rs-fMRI Data Processing Pipeline}: For each subject, we preprocess the raw rs-fMRI scans following standard neuroimaging protocols. The brain is divided into \(116\) predefined regions of interest (ROIs) using the AAL-116 atlas~\citep{tzourio2002aal}. For every ROI, we extract the blood-oxygen-level-dependent (BOLD) time series and compute Pearson correlation coefficients between all ROI pairs, yielding a symmetric \(116 \times 116\) functional connectivity (FC) matrix. 
For subject \(i\), we construct an undirected weighted graph \(G_i = (\mathcal{V}_i, \mathcal{E}_i)\), where \(|\mathcal{V}_i| = 116\) nodes correspond to ROIs, and edges \(\mathcal{E}_i\) are obtained by thresholding the FC matrix at \(\tau = 0.4\).
Each node feature is defined as the correlation profile of that ROI with all other ROIs, forming a \(116\)-dimensional feature vector. A complete visualization of the preprocessing pipeline is provided in Fig.~\ref{fig:fMRI}.




\noindent
\textbf{Continual Learning Setup}: Let $\mathcal{M}=\{M_1,\ldots,M_T\}$ denote $T$ clinical sites. Each site \(M_t\) provides its own dataset, which we split into a training set
\(\mathcal{D}_t^{\mathrm{train}}=\{(G_i^t,y_i^t)\}_{i=1}^{N_t^{\mathrm{train}}}\) of size $N_t^{\mathrm{train}}$
and a test set 
\(\mathcal{D}_t^{\mathrm{test}}=\{(G_j^t,y_j^t)\}_{j=1}^{N_t^{\mathrm{test}}}\) of size $N_t^{\mathrm{test}}$,
where \(G_i^t \in \mathcal{G}\) is the functional connectivity graph for subject \(i\) and \(y_i^t \in \mathcal{Y}\) is the diagnostic label. Tasks (sites) arrive sequentially. During training on task \(t\), the model has access 
only to \(\mathcal{D}_t^{\mathrm{train}}\) and cannot revisit 
\(\{\mathcal{D}_k^{\mathrm{train}}\}_{k<t}\). After finishing the final site \(T\), we evaluate performance on all test sets 
\(\{\mathcal{D}_1^{\mathrm{test}},\ldots,\mathcal{D}_T^{\mathrm{test}}\}\)
to assess cross-site generalization and forgetting. See Appendix~\ref{sec:description} for details.

\begin{figure}[htbp]
    \centering
    \includegraphics[width=1\linewidth]{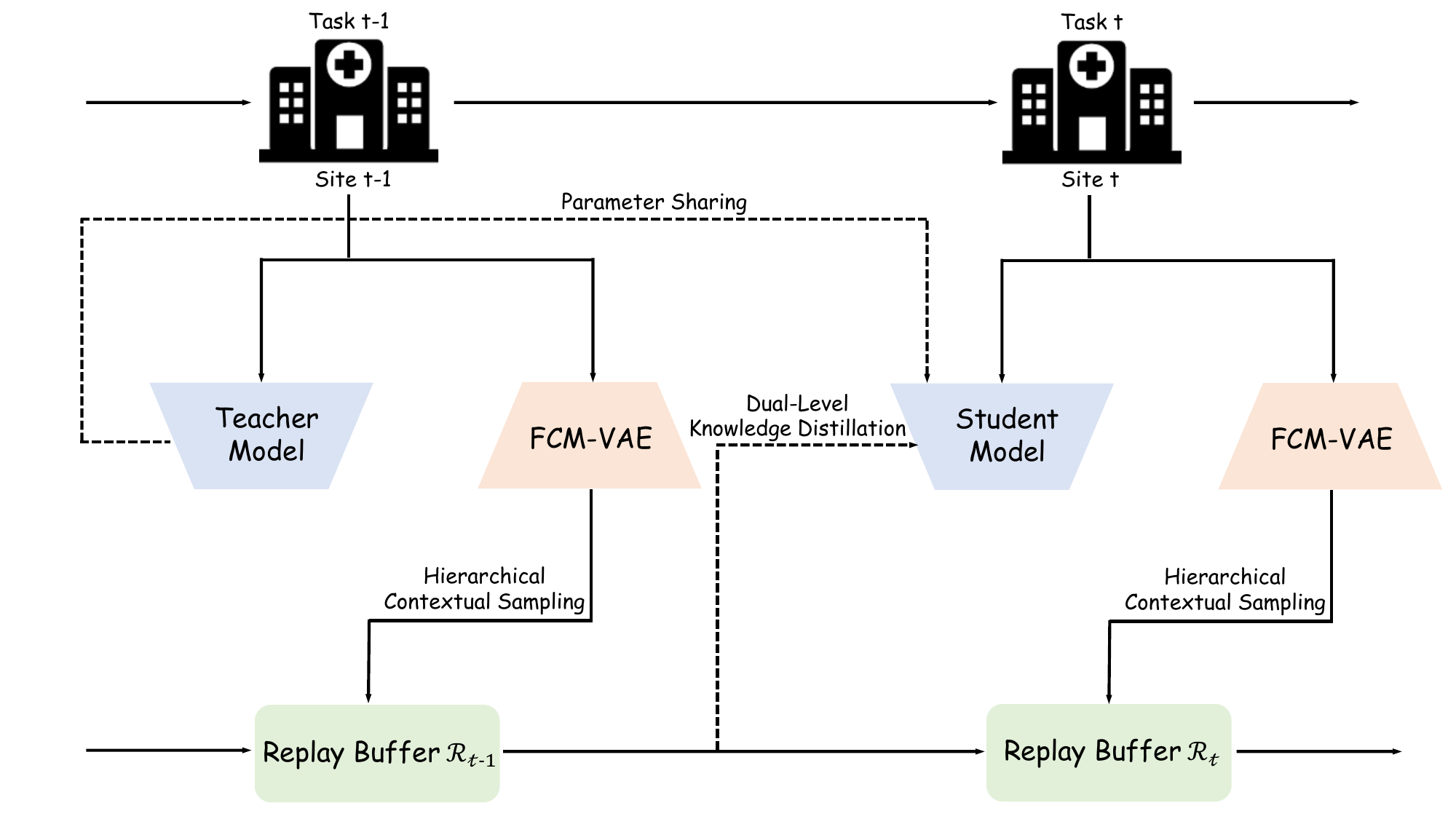}
    \caption{The continual learning workflow of FORGE across consecutive sites. At site \(t{-}1\), FCM-VAE (Section~\ref{sec:VAE}) is trained to generate functional connectivity (FC) matrices, which are stored in a replay buffer. When moving to site \(t\), a student classifier is trained to align with the previously learned model from site \(t{-}1\), matching both graph-level representations and classification logits using data from the current site and replayed samples selected via the hierarchical contextual sampling strategy (Section~\ref{sec:sampling}). After training, a new FCM-VAE is updated at site \(t\) to produce site-specific synthetic samples for future learning.}
    \label{fig:framework}
\end{figure}




\subsection{Dual-Level Knowledge Distillation Framework}

Our framework integrates generative replay into a dual-level knowledge distillation scheme. A graph classifier \(f_\theta\) is trained sequentially over tasks. For an input graph \(G\), we denote the output logits by \(h_\theta(G)\) and the corresponding predictive distribution by \(f_\theta(G)=\mathrm{softmax}(h_\theta(G))\). 


At the current task \(t_c\), we update the model parameters \(\theta\) to perform well on 
\(\mathcal{D}_{t_c}^\mathrm{train}\) while matching the behavior of the frozen teacher model 
\(f_{\theta_{t}^*}\) from earlier tasks \(t < t_c\). Here, $\theta_t^*$ is the optimal set of parameters at the end of task $t$. Formally, we minimize:
\begin{equation}
\begin{aligned}
& \mathbb{E}_{(G,y)\sim \mathcal{D}_{t_c}^\mathrm{train}} 
    \big[\ell\big(y, f_\theta(G)\big)\big] + \lambda_1 
    \sum_{t=1}^{t_c-1} 
    \mathbb{E}_{(G,y)\sim \mathcal{R}_t} 
    \big[\ell\big(y, f_\theta(G)\big)\big] \\
&\quad + \lambda_2 
    \sum_{t=1}^{t_c-1} 
    \mathbb{E}_{(G,u)\sim \mathcal{R}_t} 
    \Big[\|h_\theta(G)-u\|_2^2\Big],
\end{aligned}
\label{eq:kd}
\end{equation}
where \(\lambda_1,\lambda_2>0\), \(u \triangleq h_{\theta_t^*}(G)\) denotes the stored teacher logits, and \(\mathcal{R}_t\) is the replay buffer for task $t$. Note that replayed samples are not real FC matrices but synthetic data generated by FCM-VAE trained at each task \(t\). The first term fits the current task, the second term uses generative replay to preserve performance on earlier tasks, and the third term aligns student logits with teacher logits.


Restricting alignment to the output logits overlooks discrepancies that may arise in intermediate representations. We further introduce a graph-readout distillation mechanism that directly aligns graph-level representations between the student and the teacher. For each graph $G$, we store the teacher readout $r \triangleq \rho \big(E_{\theta_t^*}(G)\big)$, where \(E_{\theta}(\cdot)\) is the graph encoder and \(\rho(\cdot)\) is the readout function that aggregates node embeddings into a graph-level representation.
The final unified objective becomes:
\begin{equation}
\label{eq:total-pop}
\begin{aligned}
&\mathbb{E}_{(G,y)\sim\mathcal{D}_{t_c}^\mathrm{train}}\!\left[\ell\big(y,f_\theta(G)\big)\right]
+ \lambda_1 \sum_{t=1}^{t_c-1}\mathbb{E}_{(G,y)\sim\mathcal{R}_t}\!\left[\ell\big(y,f_\theta(G)\big)\right]  \\
&\quad + \lambda_2 \sum_{t=1}^{t_c-1} \mathbb{E}_{(G,u)\sim\mathcal{R}_t}
        \left[\left\|h_\theta(G)- u \right\|_2^2\right] \\
& \quad + \lambda_3 \sum_{t=1}^{t_c-1} \mathbb{E}_{(G,r)\sim\mathcal{R}_t}
        \left\|
            \rho\!\big(E_{\theta}(G)\big) - r
        \right\|_2^2,
\end{aligned}
\end{equation}
where $\lambda_{1}, \lambda_{2}, \lambda_{3} > 0$ are weighting coefficients.


For each site $M_t$, the replay buffer $\mathcal{R}_t$ stores tuples of the form $(G,\, y,\, u,\, r)$, where $G$ is a synthesized graph with label $y$,  
$u = h_{\theta_t^{*}}(G)$ denotes the teacher logits, and  
$r = \rho\big(E_{\theta_t^{*}}(G)\big)$ is the corresponding teacher graph-level representation. The teacher outputs $u$ and $r$ are computed once when the sample is added to the buffer and remain fixed thereafter, providing stable and consistent supervision during replay. See Fig.~\ref{fig:framework} for the entire framework.


\subsection{Constructing Replay Buffer with FCM-VAE}\label{sec:VAE}
\begin{figure}[htbp]
    \centering
    \includegraphics[width=0.75\linewidth]{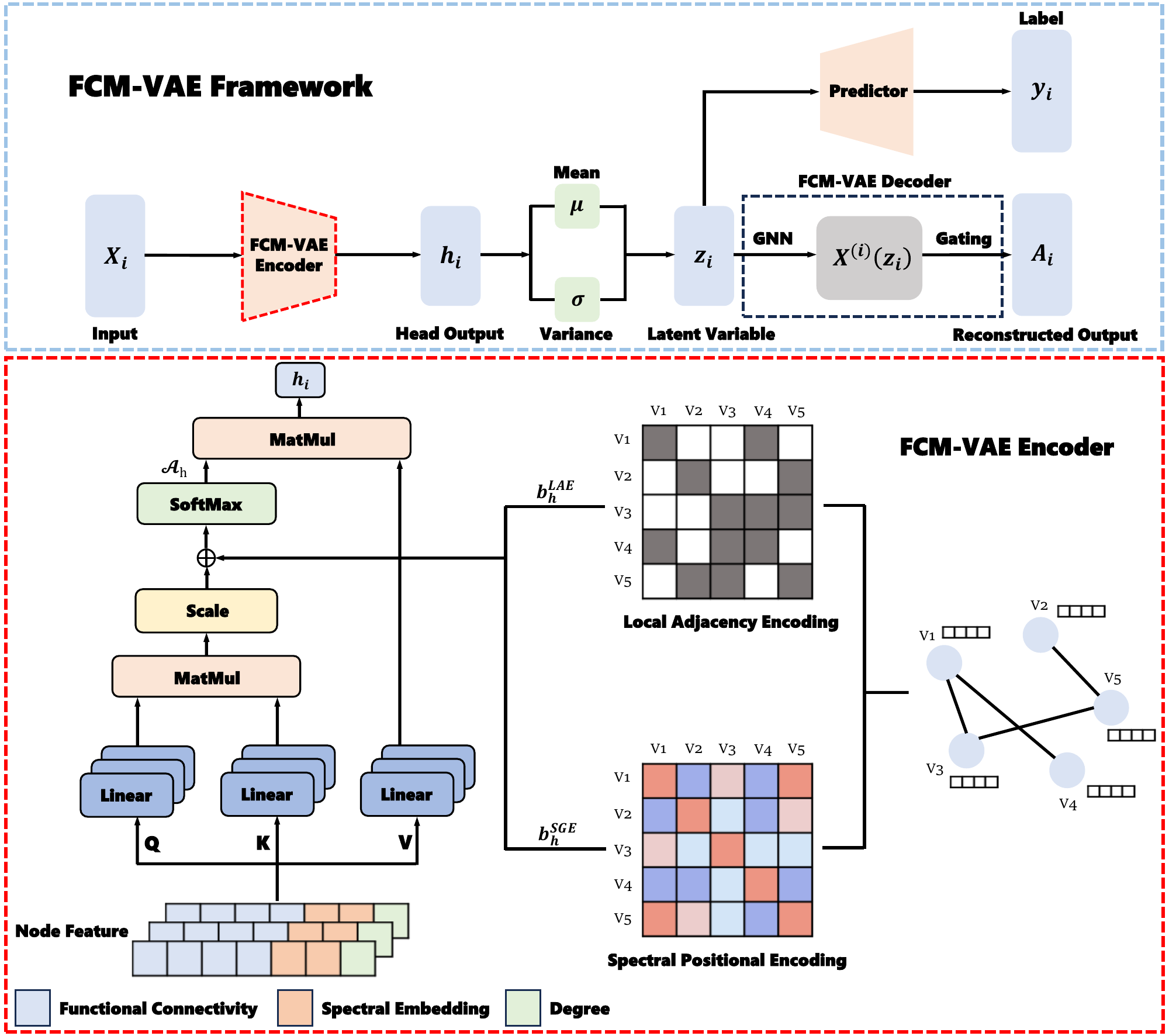}
    \caption{\textbf{Framework of FCM-VAE.}
Given an input FC graph \(G\), the encoder extracts a hidden representation \(h\), which parameterizes a diagonal Gaussian distribution with mean \(\mu\) and variance \(\sigma\), from which the latent variable \(z\) is sampled. The decoder, implemented as a gated GNN, reconstructs the adjacency matrix \(A\) and predicts the phenotype \(y\) through a linear Gaussian supervision head applied directly to \(z\). In the encoder, each node feature is formed by concatenating its functional connectivity profile, spectral embedding, and node degree. The attention layers incorporate both a local adjacency bias and a spectral positional bias, enabling the encoder to capture the intrinsic topological and spectral structure of FC networks.}
    \label{fig:placeholder}
\end{figure}

We describe the construction of our VAE-based generative model (see Fig.~\ref{fig:placeholder}) designed specifically for FC matrices.



\subsubsection{Structure-aware Graph Transformer Encoder}
To address the challenges of limited samples and high-dimensional FC matrices, we embed the topological structure and spectral geometry of each individual into the encoder for more robust and personalized representations. Concretely, we propose a Structure-aware Graph Transformer that integrates adjacency and spectral positional biases into the multi-head attention, thereby leveraging both local connectivity patterns and global geometric features. First, for each subject and their ROI node $p \in \{1, \ldots, |\mathcal{V}|\}$, we augment the initial node feature vector $x_{p}$ as:
\begin{equation}
x^{(0)}_{p}
~=~
\big[\,A^{\mathrm{ROI}}_{p},~ \phi_{p},~ \deg_{p}\,\big]
\in \mathbb{R}^{\,|\mathcal{V}| + k + 1},
\label{eq:x0}
\end{equation}
where $A^{\mathrm{ROI}}_{p} \in \mathbb{R}^{|\mathcal{V}|}$ denotes the $p$-th column of the FC matrix, $\Phi \in \mathbb{R}^{|\mathcal{V}|\times k}$ is the spectral embedding matrix formed by the top $k$ non-trivial Laplacian eigenvectors, 
and $\phi_p \in \mathbb{R}^k$ is the spectral coordinate of node $p$ in $\Phi$, and $\mathrm{deg}_{p}$ denotes the degree of node $p$, encoding its local reachability. We concatenate the $|\mathcal{V}|$ node features to form the initial feature matrix
$X^{(0)} = \big[\, x_{1}^{(0)};\, x_{2}^{(0)};\, \ldots;\, x_{|\mathcal{V}|}^{(0)} \,\big] \in \mathbb{R}^{|\mathcal{V}| \times (|\mathcal{V}| + k + 1)}$. Next, we feed $X^{(0)}$ into a multi-head self-attention layer. For each head $h$, three learnable projection matrices $W_h^Q$, $W_h^K$, and $W_h^V \in \mathbb{R}^{(|\mathcal{V}| + k + 1) \times d_h}$ are used to obtain $Q_h$, $K_h$, and $V_h \in \mathbb{R}^{|V| \times d_h}$, respectively, where $d_h=(|V|+k+1)/H$ and $H$ is the number of heads~\citep{ying2021transformersreallyperformbad}.



Each head further incorporates a local adjacency bias and a spectral positional bias into the attention scores. \emph{Local Adjacency Encoding} leverages the adjacency matrix to enforce local topological constraints and guide attention towards neighboring nodes. The attention bias is defined as:
\begin{equation}
b^{\mathrm{LAE}}_{h}~=~\tilde A
~=~A^{\mathrm{adj}}+I~\in~\mathbb{R}^{|\mathcal{V}| \times |\mathcal{V}|},
\end{equation}
where $A^{\mathrm{adj}} \in \mathbb{R}^{|\mathcal{V}| \times |\mathcal{V}|}$ is the binary adjacency matrix and 
$I$ adds self-loops so that each node includes itself when computing attention. 


\emph{Spectral positional encoding} focuses on capturing global geometric information and allows nodes with similar spectral coordinates to attend more closely to each other. The attention bias is defined as:
\begin{equation}
b^{\mathrm{SGE}}_{h}
=
\frac{\big(\Phi W^{\phi}_h\big)\big(\Phi W^{\phi}_h\big)^{\!\top}}{\sqrt{d_h}}
~\in~ \mathbb{R}^{|\mathcal{V}| \times |\mathcal{V}|},
\end{equation}
where $W^{\phi}_h \in \mathbb{R}^{k \times d_h}$ is a learnable projection that maps the spectral embedding into the $d_h$ dimensional subspace of head $h$. 


By integrating both local adjacency encoding and spectral positional encoding as additive biases, we obtain the final attention score:
\begin{equation}
  \mathcal{H}_{h}
  =
  \frac{Q_{h}K_{h}^{\!\top}}{\sqrt{d_h}}
  + \tilde A
  + \frac{\big(\Phi W^{\phi}_h\big)\big(\Phi W^{\phi}_h\big)^{\!\top}}{\sqrt{d_h}} ~\in~ \mathbb{R}^{|\mathcal{V}| \times |\mathcal{V}|}.
  \label{eq:attn-score}
\end{equation}

Using additive biases preserves the expressiveness of full-graph attention. As a result, the proposed attention mechanism remains robust and expressive even in small-sample scenarios with complex graph structures.


From the attention score matrix, we obtain the attention weights $\mathcal{A}_{h} = \mathrm{softmax}(\mathcal{H}_{h}), $ and head outputs $H_{h} = \mathcal{A}_{h} V_{h}$. Multi-head outputs are concatenated and passed through feed-forward and normalization layers for $L$ layers to yield node representations $\{x^{(L)}_{p}\}_{p=1}^{|\mathcal{V}|}$. The graph-level embedding is obtained through a permutation-invariant readout function $h = \mathrm{Readout}\!\left(\tfrac{1}{|\mathcal{V}|}\sum_{p=1}^{|\mathcal{V}|} x^{(L)}_{p}\right)$, where $\mathrm{Readout}(\cdot)$ applies mean pooling over node representations followed by a linear projection.

We then employ $h$ to parameterize the approximate posterior of the latent variable as a diagonal-covariance Gaussian distribution, following the design principle of VAE:
\begin{equation}
q_\phi(z \mid h)
= \mathcal{N}\!\big(\mu_\phi(h), \mathrm{diag}[\sigma_\phi^2(h)]\big),\qquad z \in \mathbb{R}^{d_z}
\label{eq:posterior}
\end{equation}
where $\mu_\phi(h), \sigma^2_\phi(h) \in \mathbb{R}^{d_z}$ are encoder outputs that parameterize the mean and variance of the Gaussian posterior. This design preserves permutation invariance at the graph level while keeping the posterior parameterization simple.



\subsubsection{Low‑rank Decoder with Reachability Gating}

On the decoder side, we reconstruct the functional connectivity matrix from the low-dimensional latent variable $z$. We assume that the reconstructed edge entries are conditionally independent given $z \in \mathbb{R}^{d_z}$~\citep{LIU2021118750}. Accordingly, the likelihood of the reconstructed graph $\widehat{A}$ factorizes as:
\begin{equation}
p_\psi(L(\widehat{A}) = a \mid z)
=
\prod_{e=1}^{|\mathcal{V}|(|\mathcal{V}|-1)/2}
p_\psi \big(\widehat{A}_{e} = a_{e} \mid z\big),
\label{eq:decoder_likelihood}
\end{equation}
where $p_\psi(\widehat{A}_{e} \mid z)$ denotes the generative model for edge $e$, and we impose a Gaussian prior $p(z)=\mathcal{N}(0,I_{d_z})$.

To model connectivity strengths with positive support, we first transform each Pearson correlation entry $r_e \in [-1,1]$ into a nonnegative intensity. Specifically, we apply the Fisher-$z$ transform, standardize it using training-set statistics, and then exponentiate:
\begin{equation}
s_e = \operatorname{atanh}(r_e),\quad
\tilde{s}_e = \frac{s_e-\mu_s}{\sigma_s},\quad
x_e = \exp(\tilde{s}_e),
\label{eq:corr_to_intensity}
\end{equation}
where $\mu_s$ and $\sigma_s$ are the mean and standard deviation estimated from the training data. 
In Eq.~\eqref{eq:decoder_likelihood}, we therefore let $a_e$ denote the transformed intensity $x_e$.

Given $z$, the decoder models each transformed edge intensity with a Poisson likelihood:
\begin{equation}
z \sim \mathcal{N}(0,I_{d_z}), \qquad
\widehat{A}_e \mid z \sim \mathrm{Poisson} \big(\widehat{\lambda}_e(z)\big),
\label{eq:decoder_process}
\end{equation}
where the rate is parameterized with the canonical log link,
\begin{equation}
\widehat{\lambda}_e(z)=\exp \big(\nu_e+\omega_e(z)\big),
\label{eq:lambda_decoder}
\end{equation}
where $\nu_e$ is an edge-specific baseline parameter shared across subjects, capturing the common connectivity tendency for edge $e$, while $\omega_e(z)$ models the subject-specific deviation induced by the latent representation $z$~\citep{LIU2021118750}.

For edge $e=[u,v]$, the interaction term is defined as:
\begin{equation}
\omega_e(z)=\sum_{r=1}^{R}\alpha_r\, U_{ur}(z)\,U_{vr}(z),
\label{eq:psi_interaction}
\end{equation}
where
\begin{equation}
U(z) = [U_{ur}(z)] \in \mathbb{R}^{|\mathcal{V}| \times R}, \qquad U_r(z) = \big(U_{1r}(z), \ldots, U_{|\mathcal{V}|r}(z)\big)^\top = g_r(z).
\label{eq:node_factor_matrix}
\end{equation}
Here, $U(z)$ is a decoder-generated node-factor matrix, $\alpha_r$ controls the contribution of the $r$-th latent factor, and $g_r(\cdot)$ is a nonlinear mapping parameterized by $\psi$. This factorization defines a low-rank bilinear decoder, in which each edge strength is reconstructed from the interaction of the two nodes across $R$ latent factors~\citep{hoff2002latent}. Finally, all predicted edge intensities are assembled to form the reconstructed functional connectivity matrix $\widehat{A}$.

Since FC matrices typically exhibit sparse individual-specific variations across subjects, we adopt a soft gating mechanism based on the adjacency matrix to constrain the strengths of the reconstructed edges, while avoiding spurious edge formation. We project the adjacency matrix of each subject into a gating matrix that modulates the decoder outputs:
\begin{equation}
\label{eq:gating}
G=\sigma\!\big(A^{\mathrm{adj}} + \mathrm{logit}(\varepsilon)\big),
\qquad
A^{\mathrm{pred}} = G \odot \widehat{A}.
\end{equation}
Here, $\sigma(\cdot)$ denotes the element-wise sigmoid function, $\odot$ is the Hadamard product, and $\varepsilon$ is a small constant (e.g., $10^{-2}$). 


\subsubsection{Linear Gaussian Supervised Head and Conditional Generation}
To jointly reconstruct the graph and predict the phenotype $y$, we attach a linear–Gaussian supervision module on $z$. We posit a linear Gaussian relationship between the phenotype and the latent factors, leading to the following formulation:
\begin{equation}
    p(y \mid z) = \mathcal{N}\big(\beta_0 + \bm{\beta}^{\top} z, \sigma_y^2\big),
  \label{eq:supervised}
\end{equation}
Here, $\beta_0$ and $\beta$ are learnable parameters, and $\sigma_y^2$ denotes the observation noise variance. The linear Gaussian supervision head introduces minimal inductive bias by directly linking the label $y$ to the latent variable, improving interpretability and calibration of the latent space. At inference, prediction is obtained from the posterior mean as $\hat{y} = \beta_0 + \beta^\top \mu_\phi(h)$, where $\mu_\phi(h)$ is the mean of the latent posterior given $h$.

\subsubsection{Overall Training Objective}
FCM-VAE is trained under a supervised evidence lower bound that jointly optimizes graph reconstruction and class prediction. The training objective is formulated as:
\begin{equation}
\mathcal{L} = \mathbb{E}_{q_\phi}\!\big[\log p_\psi(A^{\mathrm{ROI}} \mid z)\big] + \mathbb{E}_{q_\phi}\!\big[\log p_\theta(y \mid z)\big] - \beta \, D_{\mathrm{KL}}\!\left(q_\phi(z \mid \cdot) \,\|\, p(z)\right)
\label{eq:elbo}
\end{equation}
where $A^{\mathrm{ROI}}$ and $y$ denote the observed FC matrix and phenotype label of the current subject, and $\beta>1$ serves as a regularization strength that encourages a compact latent representation and mitigates overfitting. The first term enforces accurate reconstruction between
\(A^{\text{pred}}\) and \(A^{\text{ROI}}\). 
The second term is a supervised likelihood that shapes the latent space so that \(z\) captures information predictive of the phenotype \(y\). The third term encourages the variational posterior \(q_{\phi}(z \mid \cdot)\) to remain close to the prior distribution \(p(z)\), chosen as a standard Gaussian as in the conventional VAE.

\subsection{Hierarchical Contextual Thompson Sampling}\label{sec:sampling}
To make efficient use of a limited replay memory, we propose a Hierarchical Contextual Thompson Sampling (HCTS) framework that adaptively allocates replay capacity across hospital sites (site level) and selects the most informative samples within each site (sample level).

\subsubsection{Site-Level Sampling}
We assume a total replay capacity of \(K\) samples shared across all sites. The goal is to allocate \(k_i\) samples to each site \(M_i\), such that \(\sum_i k_i = K\). For each site, we define a context vector \(\phi_i = [\mathrm{Acc}_i, \mathrm{Forget}_i]\), where \(\mathrm{Acc}_i\) denotes current accuracy and \(\mathrm{Forget}_i\) measures performance decay. The expected replay utility is modeled as:
\begin{equation}
r_i = \phi_i^\top \mathbf{w} + \epsilon_i, \quad \epsilon_i \sim \mathcal{N}(0, \sigma^2),
\end{equation}
where \(\mathbf{w}\) is the latent weight vector with posterior \(p(\mathbf{w}\mid\mathcal{D})\). At each training round, Thompson Sampling draws a posterior sample \(\tilde{\mathbf{w}}\) and estimates the utility:
\begin{equation}
\tilde{r}_i = \phi_i^\top \tilde{\mathbf{w}}.
\end{equation}

The replay quota for each site is then determined by a normalized softmax allocation:
\begin{equation}
k_i = K \cdot \frac{\exp(\tilde{r}_i)}{\sum_j \exp(\tilde{r}_j)},
\end{equation}
ensuring \(\sum_i k_i = K\). This mechanism prioritizes sites with higher uncertainty or forgetting, enhancing global stability.

\subsubsection{Sample-Level Sampling}
Given the site-specific quota \(k_i\), we further identify the most beneficial samples from each buffer \(\mathcal{R}_i\). Each sample \(u \in \mathcal{R}_i\) is described by a context vector \(\psi_u = [\mathrm{margin}_u, \mathrm{closeness}_u]\), capturing uncertainty and representativeness. The expected utility is modeled as:
\begin{equation}
r_u = \psi_u^\top \mathbf{v} + \epsilon_u, \quad \epsilon_u \sim \mathcal{N}(0, \sigma^2),
\end{equation}
where \(\mathbf{v}\) denotes the sample-level relevance weights. Thompson Sampling draws \(\tilde{\mathbf{v}}\) from the posterior to compute:
\begin{equation}
\tilde{r}_u = \psi_u^\top \tilde{\mathbf{v}}.
\end{equation}

The top-\(k_i\) samples with the highest \(\tilde{r}_u\) values are selected with a greedy farthest-first traversal in the readout embedding space 
$\{\rho(E_{\theta}(G)) \mid G \in \mathcal{C}_i \}$, 
thereby improving coverage and reducing redundancy. 
The selection objective is formulated as:
\begin{equation}
\label{eq:ffs-obj}
\mathcal{S}_i
= 
\arg\max_{S \subseteq \mathcal{C}_i :\, |S| = k_i}
\;
\min_{\substack{G,\,G' \in S \\ G \neq G'}}
\big\|
    \rho(E_{\theta}(G))
    -
    \rho(E_{\theta}(G'))
\big\|_2.
\end{equation}

This hierarchical design ensures replay focuses both on \emph{where} forgetting occurs (site level) and \emph{which} samples are most valuable for continual adaptation (sample level). We refer interested readers to Appendix~\ref{sec:HCTS} for details.

\section{Experiments and Results}
\label{Experiments and Results}

\subsection{The Datasets and Preprocessing}
To rigorously assess the proposed FORGE framework, we conduct experiments on three widely used, large-scale, multi-site fMRI datasets: the Autism Brain Imaging Data Exchange I (ABIDE-I)~\citep{ASD}, the REST-meta-MDD dataset~\citep{MDD}, and the Bipolar and Schizophrenia Network for Intermediate Phenotypes (BSNIP)~\citep{BSNIP}. The ABIDE-I dataset consists of rs-fMRI scans from 1099 participants across 17 independent sites. To ensure statistical reliability, we focus on the four largest sites in the ABIDE-I dataset: Site 6 (184 samples), Site 14 (99), Site 15 (145), and Site 16 (101). For the REST-meta-MDD dataset, the largest rs-fMRI resource for major depressive disorder (MDD), we follow the quality-control criteria~\citep{10680255} and select the five largest sites: Site 1 (146 samples), Site 6 (96), Site 10 (93), Site 14 (470), and Site 15 (144). The BSNIP contains 1,142 individuals, comprising 640 healthy controls and 502 participants diagnosed with schizophrenia. Similarly, we retain the five largest sites: Hartford (297 samples), Chicago (239), Dallas (197), Georgia (152), and Baltimore (120). All three datasets were preprocessed using a standard rs-fMRI pipeline implemented with the Statistical Parametric Mapping (SPM12) software and the DPABI toolbox~\cite{yan2016dpabi}; details are provided in Appendix~\ref{sec:preprocessing}.



\subsection{Evaluation of FCM-VAE with Data Augmentation}
\begin{table}[htbp]
\centering
\caption{FC classification results with data augmentation using different graph generation methods. Columns report ACC, PRE, REC, and F1. Bold numbers denote the best results, and underlined numbers indicate the second-best.}
\label{tab:augmentation}
\resizebox{\textwidth}{!}{%
\begin{tabular}{
  l l
  S[table-format=1.3] S[table-format=1.3] S[table-format=1.3] S[table-format=1.3]
  S[table-format=1.3] S[table-format=1.3] S[table-format=1.3] S[table-format=1.3]
  S[table-format=1.3] S[table-format=1.3] S[table-format=1.3] S[table-format=1.3]
}
\toprule
\multirow{2}{*}{\textbf{Classifier}} & \multirow{2}{*}{\textbf{Gen. Method}} &
\multicolumn{4}{c}{\textbf{ASD}} & \multicolumn{4}{c}{\textbf{BSNIP}} & \multicolumn{4}{c}{\textbf{MDD}} \\
\cmidrule(lr){3-6}\cmidrule(lr){7-10}\cmidrule(lr){11-14}
& & \textbf{ACC} & \textbf{PRE} & \textbf{REC} & \textbf{F1}
  & \textbf{ACC} & \textbf{PRE} & \textbf{REC} & \textbf{F1}
  & \textbf{ACC} & \textbf{PRE} & \textbf{REC} & \textbf{F1} \\
\midrule

\multirow{5}{*}{\textbf{GCN}}
& No Augmentation      & \m{0.762}{0.748}{0.772}{0.758} & \m{0.688}{0.733}{0.456}{0.502} & \m{0.756}{\topone{0.733}}{0.699}{0.700} \\
& ReGate               & \m{0.724}{0.688}{\toptwo{0.806}}{0.740} & \m{0.639}{0.618}{0.318}{0.361} & \m{0.724}{0.701}{0.725}{0.676} \\
& BrainNetGan          & \m{0.758}{0.757}{0.752}{0.747} & \m{0.686}{\toptwo{0.747}}{\toptwo{0.458}}{0.488} & \m{0.748}{\toptwo{0.707}}{0.745}{0.712} \\
& GR-SPD-GAN           & \m{\toptwo{0.780}}{\toptwo{0.761}}{\topone{0.812}}{\topone{0.780}} & \m{\toptwo{0.703}}{\topone{0.785}}{0.452}{\toptwo{0.513}} & \m{\toptwo{0.767}}{0.666}{\topone{0.866}}{\topone{0.745}} \\
& \textbf{FCM-VAE (Ours)} & \m{\topone{0.783}}{\topone{0.776}}{0.775}{\toptwo{0.771}} & \m{\topone{0.727}}{0.727}{\topone{0.547}}{\topone{0.597}} & \m{\topone{0.787}}{\topone{0.733}}{\toptwo{0.784}}{\toptwo{0.735}} \\
\midrule

\end{tabular}%
}
\end{table}
We begin by quantitatively assessing the quality of synthetic FC matrices generated by FCM-VAE. Specifically, for each site in each dataset, we split the data into 70\% training and 30\% testing, generate a synthetic FC graph for every training sample, and augment the training set with an equal number of synthetic samples. A GCN classifier~\citep{kipf2017semisupervisedclassificationgraphconvolutional} is trained on the augmented data and evaluated on the held-out test split. We compare FCM-VAE against three state-of-the-art brain network generative models, including ReGate~\citep{LIU2021118750}, BrainNetGAN~\citep{li2021brainnetgandataaugmentationbrain}, and GR-SPD-GAN~\citep{tan2022graphregularizedmanifoldawareconditionalwasserstein}, and also report results without augmentation. Performance is measured using accuracy, precision, recall, and F1-score. As shown in Table~\ref{tab:augmentation}, FCM-VAE consistently yields the best or second-best performance across all datasets, particularly in classification accuracy, with GR-SPD-GAN ranking second overall.

Additional results using other classifiers, such as GAT~\citep{velivckovic2017graph}, GraphSAGE~\citep{hamilton2018inductiverepresentationlearninglarge}, GAE~\citep{kipf2016variationalgraphautoencoders}, and GraphTransformer~\citep{dwivedi2021generalizationtransformernetworksgraphs}, as well as visualizations of synthesized FC matrices produced by different approaches (e.g., FCM-VAE and GR-SPD-GAN), are provided in Appendix~\ref{sec:FCM_VAE_more}. 


\subsection{Evaluation of FORGE}
In this section, we systematically evaluate the proposed continual learning framework FORGE under a unified experimental protocol. To comprehensively assess its adaptability and resilience to catastrophic forgetting in multi‑task and cross‑site settings, four representative categories of continual learning methods are adopted as baselines for comparison: regularization-based methods (EWC~\citep{Kirkpatrick_2017}, SI~\citep{zenke2017continuallearningsynapticintelligence}), parameter isolation approaches (PackNet~\citep{mallya2018packnet}), rehearsal-based strategies (ER~\citep{rolnick2019experiencereplaycontinuallearning}), and knowledge distillation methods (LwF~\citep{li2017learning}, DER~\citep{buzzega2020dark}, DER++~\citep{buzzega2020dark}). In addition, we also consider three modern frameworks specifically designed for continual learning on graph classification tasks, including TWP~\citep{liu2020overcomingcatastrophicforgettinggraph}, UGCL~\citep{hoang2023universalgraphcontinuallearning}, and PDGNNs~\citep{Zhang_2024}. According to the results in Table~\ref{tab:three-datasets-reordered}, GCN achieves the highest average accuracy among all candidate classifiers and is therefore adopted as the backbone for all continual learning frameworks. For each site, the data are split into a 7:3 ratio for training and testing, and we fix the replay buffer size to 256 for ER, DER, DER++, and FORGE.

To evaluate continual learning performance, we adopt two standard metrics that respectively reflect stability and plasticity: Average Anytime Accuracy (AAA) and Forgetting Rate (FOR). AAA measures the overall generalization ability across sequential tasks, while FOR quantifies the degree of catastrophic forgetting. Formally,
\begin{equation}
\mathrm{AAA} = \frac{1}{T}\sum_{t=1}^{T}\frac{1}{t}\sum_{i=1}^{t}\mathrm{ACC}_{t,i}, \qquad
\mathrm{FOR} = \frac{1}{T}\sum_{i=1}^{T}\big(\max_{t \in \{1,\ldots,T-1\}}\mathrm{ACC}_{t,i} - \mathrm{ACC}_{T,i}\big).
\label{eq:aaa_for}
\end{equation}
where \(T\) is the total number of tasks, \(\mathrm{ACC}_{t,i}\) denotes the accuracy on task \(i\) after completing task \(t\), and \(\mathrm{ACC}_{T,i}\) is the final accuracy on task \(i\) after learning all tasks. To ensure fair comparison, we evaluate each method under all task orders and report the averaged performance across sequences.
\begin{table}[htbp]
\centering
\footnotesize
\setlength{\tabcolsep}{5pt}
\renewcommand{\arraystretch}{1.05}
\caption{Performance of different continual learning methods on ASD, BSNIP, and MDD.}
\label{tab:cl_means_all}
\begin{tabular}{l *{6}{S[table-format=1.4]}}
\toprule
& \multicolumn{2}{c}{\textbf{ASD}} & \multicolumn{2}{c}{\textbf{BSNIP}} & \multicolumn{2}{c}{\textbf{MDD}} \\
\cmidrule(lr){2-3}\cmidrule(lr){4-5}\cmidrule(lr){6-7}
\textbf{CL Method} & {\textbf{AAA}$\uparrow$} & {\textbf{FOR}$\downarrow$} & {\textbf{AAA}$\uparrow$} & {\textbf{FOR}$\downarrow$} & {\textbf{AAA}$\uparrow$} & {\textbf{FOR}$\downarrow$} \\
\midrule
Base            & 0.6074 & 0.2757 & 0.5512 & 0.2296 & 0.5929 & 0.2770 \\
DER             & 0.6627 & 0.2018 & 0.6320 & 0.1254 & 0.6655 & 0.1753 \\
DER++           & 0.6719 & 0.1893 & 0.6225 & 0.1384 & 0.6950 & 0.1471 \\
ER              & 0.6369 & 0.2397 & 0.5967 & 0.1702 & 0.6482 & 0.1757 \\
EWC             & 0.5963 & 0.1925 & 0.5688 & 0.1897 & 0.6074 & 0.2018 \\
LwF             & 0.6464 & 0.2222 & 0.6218 & 0.1423 & 0.6388 & 0.1858 \\
PackNet         & 0.6632 & 0.2030 & 0.6178 & 0.1355 & 0.6235 & 0.1864 \\
SI              & 0.6684 & 0.1746 & 0.6553 & 0.1365 & 0.6475 & 0.1757 \\
TWP             & 0.6735 & 0.1659 & 0.6391 & 0.1472 & 0.6777 & 0.1795 \\
UGCL            & 0.6795 & 0.1589 & 0.6398 & 0.1439 & 0.6915 & 0.1613 \\
PDGNNs          & 0.7113 & 0.1313 & 0.6588 & 0.1281 & 0.7050 & 0.1392 \\
\textbf{FORGE (Ours)} & {\bfseries 0.7300} & {\bfseries 0.1160} & {\bfseries 0.6850} & {\bfseries 0.1010} & {\bfseries 0.7380} & {\bfseries 0.1050} \\
\bottomrule
\end{tabular}
\end{table}
\begin{figure}[htbp]
    \centering
    \includegraphics[width=1\linewidth]{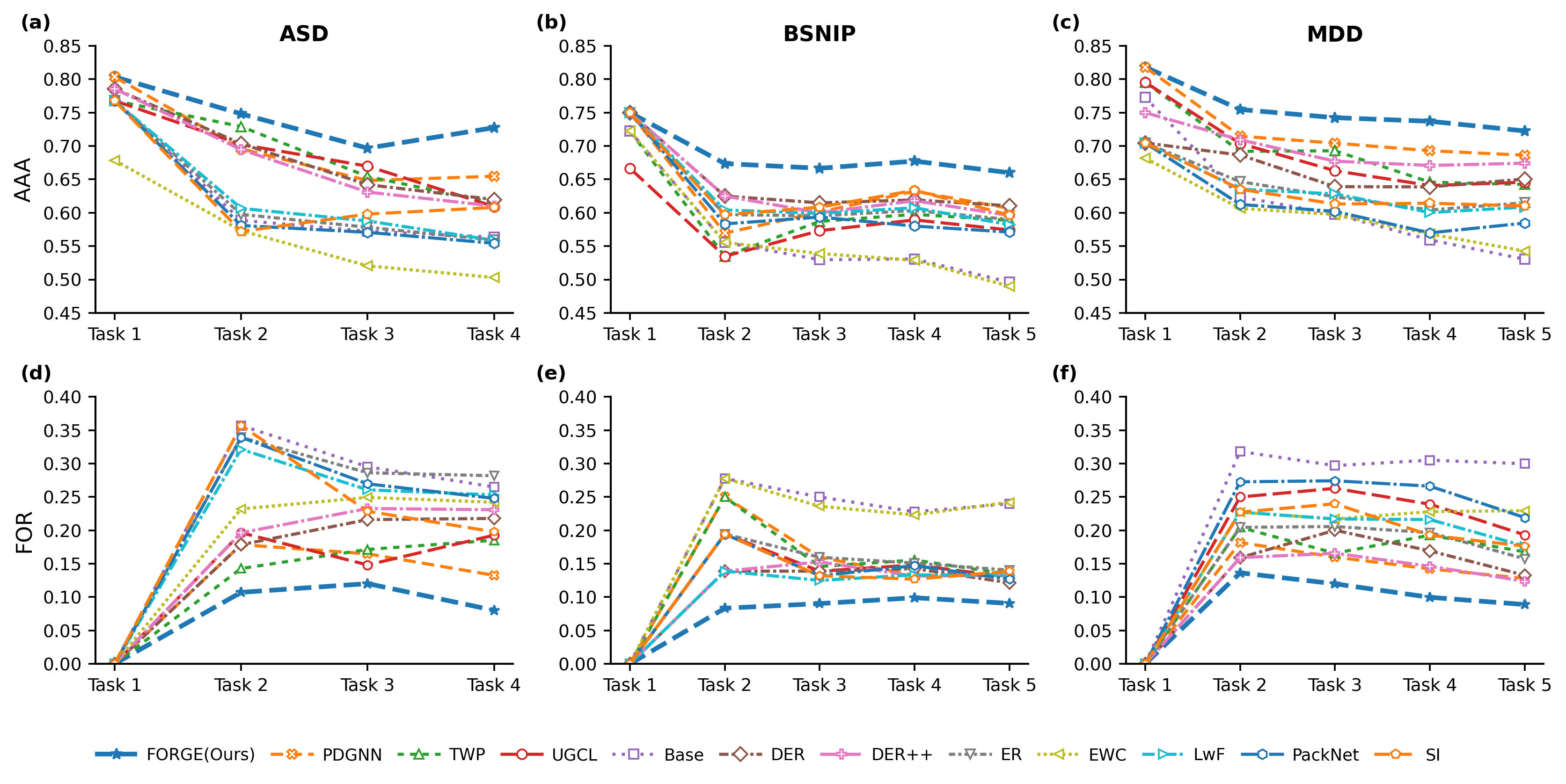}
    \caption{Visualization of the averaged AAA and FOR across sequential tasks under a fixed task order on ASD, BSNIP, and MDD.}
    \label{fig:clresult}
\end{figure}

From the results in Table~\ref{tab:cl_means_all}, FORGE consistently achieves the best balance between stability and plasticity across all three benchmarks, obtaining the highest AAA scores and the lowest FOR values. Relative to the strongest baseline on each dataset, FORGE improves AAA by 2.7\%, 3.9\%, and 4.7\% on ASD, BSNIP, and MDD, respectively, and reduces forgetting by 11.5\%, 21.1\%, and 24.5\%. While PDGNNs stands as the second-best method overall, highlighting the benefits of topology design for mitigating catastrophic forgetting in graph-based continual learning. Meanwhile, knowledge distillation–based approaches such as DER and DER++ also exhibit competitive performance, demonstrating the strong potential of distillation strategies in preserving both stability and plasticity.

Furthermore, we visualize the averaged AAA and FOR across all task learning stages, as illustrated in Figure~\ref{fig:clresult}. It can be observed that FORGE maintains superior accuracy throughout the continual learning process. In contrast, other methods suffer from noticeable performance drops, particularly in later stages, revealing their limited ability to retain previous knowledge and adapt stably over time.

\subsection{Ablation Study and Sensitivity Analysis}
\begin{table}[htbp]
\centering
\footnotesize
\setlength{\tabcolsep}{5pt}
\renewcommand{\arraystretch}{1.05}
\caption{Ablation study assessing the contributions of Graph Knowledge Distillation (GKD), Logits Knowledge Distillation (LKD), and Hierarchical Contextual Thompson Sampling (HCTS) to the overall performance of FORGE.}
\label{tab:ablation_all}
\begin{tabular}{l *{6}{S[table-format=1.4]}}
\toprule
& \multicolumn{2}{c}{\textbf{ASD}} & \multicolumn{2}{c}{\textbf{BSNIP}} & \multicolumn{2}{c}{\textbf{MDD}} \\
\cmidrule(lr){2-3}\cmidrule(lr){4-5}\cmidrule(lr){6-7}
\textbf{Method} & {\textbf{AAA}$\uparrow$} & {\textbf{FOR}$\downarrow$} & {\textbf{AAA}$\uparrow$} & {\textbf{FOR}$\downarrow$} & {\textbf{AAA}$\uparrow$} & {\textbf{FOR}$\downarrow$} \\
\midrule
Full Framework & 0.7300 & 0.1160 & 0.6850 & 0.1010 & 0.7380 & 0.1050 \\
w/o GKD        & 0.6830 & 0.1424 & 0.6374 & 0.1313 & 0.6913 & 0.1410 \\
w/o LKD        & 0.6807 & 0.1508 & 0.6380 & 0.1252 & 0.6954 & 0.1331 \\
w/o HCTS       & 0.6759 & 0.1629 & 0.6342 & 0.1365 & 0.6848 & 0.1439 \\
\bottomrule
\end{tabular}
\end{table}
To further investigate the contribution of each component in our framework, we conduct an ablation study as summarized in Table \ref{tab:ablation_all}. Removing different components from FORGE leads to distinct performance degradations.
The absence of GKD results in higher forgetting and lower AAA, while removing LKD further weakens task-specific retention.
Eliminating the HCTS causes the most severe drop, yielding the highest forgetting across all datasets.


\begin{table}[htbp]
\centering
\footnotesize
\setlength{\tabcolsep}{5pt}
\renewcommand{\arraystretch}{1.05}
\caption{Performance of continual learning methods under different buffer sizes on the ASD dataset.}
\label{tab:buffer_size_ablation}
\begin{tabular}{c *{8}{S[table-format=1.4]}}
\toprule
& \multicolumn{2}{c}{\textbf{FORGE (Ours)}} 
& \multicolumn{2}{c}{\textbf{ER}}
& \multicolumn{2}{c}{\textbf{DER}}
& \multicolumn{2}{c}{\textbf{DER++}} \\
\cmidrule(lr){2-3}
\cmidrule(lr){4-5}
\cmidrule(lr){6-7}
\cmidrule(lr){8-9}
\textbf{Buffer Size} 
& {\textbf{AAA}$\uparrow$} & {\textbf{FOR}$\downarrow$}
& {\textbf{AAA}$\uparrow$} & {\textbf{FOR}$\downarrow$}
& {\textbf{AAA}$\uparrow$} & {\textbf{FOR}$\downarrow$}
& {\textbf{AAA}$\uparrow$} & {\textbf{FOR}$\downarrow$} \\
\midrule
128 & 0.6853 & 0.1570 & 0.6008 & 0.2746 & 0.6278 & 0.2214 & 0.6265 & 0.2285 \\
256 & 0.7300 & 0.1160 & 0.6370 & 0.2400 & 0.6630 & 0.2020 & 0.6720 & 0.1890 \\
384 & 0.7194 & 0.1285 & 0.6424 & 0.2154 & 0.6577 & 0.2166 & 0.6693 & 0.2096 \\
\bottomrule
\end{tabular}
\end{table}

To assess the impact of memory capacity, we evaluate FORGE and competing methods using buffer under different buffer sizes, as reported in Table~\ref{tab:buffer_size_ablation}. FORGE consistently achieves the best performance across all settings, with the advantage most evident in the low-buffer regime. This robustness stems from HCTS, which effectively selects more informative and diverse synthetic samples to maximize the utility of the constrained buffer. 
Interestingly, performance declines slightly as the buffer size increases. This may be because a larger buffer size introduces more redundant or less informative generated samples.


\begin{table}[htbp]
\centering
\footnotesize
\setlength{\tabcolsep}{5pt}
\renewcommand{\arraystretch}{1.05}
\caption{Sensitivity to the FC threshold $\tau$ on ASD.}
\label{tab:tau_sensitivity_all_methods}
\begin{tabular}{l *{8}{S[table-format=1.3]}}
\toprule
& \multicolumn{2}{c}{\textbf{Base}} 
& \multicolumn{2}{c}{\textbf{DER++}} 
& \multicolumn{2}{c}{\textbf{PDGNN}} 
& \multicolumn{2}{c}{\textbf{FORGE (Ours)}} \\
\cmidrule(lr){2-3}\cmidrule(lr){4-5}\cmidrule(lr){6-7}\cmidrule(lr){8-9}
$\tau$ 
& {\textbf{AAA}$\uparrow$} & {\textbf{FOR}$\downarrow$}
& {\textbf{AAA}$\uparrow$} & {\textbf{FOR}$\downarrow$}
& {\textbf{AAA}$\uparrow$} & {\textbf{FOR}$\downarrow$}
& {\textbf{AAA}$\uparrow$} & {\textbf{FOR}$\downarrow$} \\
\midrule
0.2 & 0.595 & 0.306 & 0.655 & 0.200 & 0.709 & 0.123 & {\bfseries 0.727} & {\bfseries 0.109} \\
0.3 & 0.589 & 0.273 & 0.670 & 0.197 & 0.707 & 0.125 & {\bfseries 0.721} & {\bfseries 0.114} \\
0.4 & 0.607 & 0.276 & 0.672 & 0.189 & 0.711 & 0.131 & {\bfseries 0.730} & {\bfseries 0.116} \\
0.5 & 0.613 & 0.293 & 0.680 & 0.213 & 0.716 & 0.143 & {\bfseries 0.735} & {\bfseries 0.120} \\
\bottomrule
\end{tabular}
\end{table}

We further examine the sensitivity of our framework to different threshold values of $\tau$ in graph construction. As shown in Table~\ref{tab:tau_sensitivity_all_methods}, \textsc{FORGE} demonstrates strong robustness to different $\tau$ values, consistently yielding the best AAA and the lowest FOR. This suggests that the observed gains are stable rather than tied to a particular threshold setting. Additional ablation results are provided in Appendix~\ref{sec:additional_ablation}.
\section{Conclusion}

We introduced FORGE, the first continual learning framework for fMRI-based brain disorder diagnosis. At its core, FCM-VAE is a structure-aware generative model that synthesizes high-quality functional connectivity matrices, providing a viable solution for privacy-preserving replay in clinical environments. FORGE further stabilizes learning through dual-level distillation and adaptive sampling, ensuring knowledge retention as new sites arrive. Experiments on three large-scale neuroimaging datasets demonstrate consistent improvements in accuracy and substantial reductions in forgetting compared with state-of-the-art approaches.


\bibliography{biblio.bib}

@String(AAAI = {AAAI})

@article{plitt2015functional,
  title={Functional connectivity classification of autism identifies highly predictive brain features but falls short of biomarker standards},
  author={Plitt, Mark and Barnes, Kelly Anne and Martin, Alex},
  journal={NeuroImage: Clinical},
  volume={7},
  pages={359--366},
  year={2015},
  publisher={Elsevier}
}

@article{kawahara2017brainnetcnn,
  title={BrainNetCNN: Convolutional neural networks for brain networks; towards predicting neurodevelopment},
  author={Kawahara, Jeremy and Brown, Colin J and Miller, Steven P and Booth, Brian G and Chau, Vann and Grunau, Ruth E and Zwicker, Jill G and Hamarneh, Ghassan},
  journal={NeuroImage},
  volume={146},
  pages={1038--1049},
  year={2017},
  publisher={Elsevier}
}

@article{gallo2023functional,
  title={Functional connectivity signatures of major depressive disorder: machine learning analysis of two multicenter neuroimaging studies},
  author={Gallo, Selene and El-Gazzar, Ahmed and Zhutovsky, Paul and Thomas, Rajat M and Javaheripour, Nooshin and Li, Meng and Bartova, Lucie and Bathula, Deepti and Dannlowski, Udo and Davey, Christopher and others},
  journal={Molecular Psychiatry},
  volume={28},
  number={7},
  pages={3013--3022},
  year={2023},
  publisher={Nature Publishing Group UK London}
}

@article{parisi2019continual,
  title={Continual lifelong learning with neural networks: A review},
  author={Parisi, German I and Kemker, Ronald and Part, Jose L and Kanan, Christopher and Wermter, Stefan},
  journal={Neural networks},
  volume={113},
  pages={54--71},
  year={2019},
  publisher={Elsevier}
}

@article{van2019three,
  title={Three scenarios for continual learning},
  author={Van de Ven, Gido M and Tolias, Andreas S},
  journal={arXiv preprint arXiv:1904.07734},
  year={2019}
}

@article{amrollahi2022leveraging,
  title={Leveraging clinical data across healthcare institutions for continual learning of predictive risk models},
  author={Amrollahi, Fatemeh and Shashikumar, Supreeth P and Holder, Andre L and Nemati, Shamim},
  journal={Scientific reports},
  volume={12},
  number={1},
  pages={8380},
  year={2022},
  publisher={Nature Publishing Group UK London}
}

@article{zeng2019multi,
  title={Multi-site fMRI analysis using privacy-preserving federated learning and domain adaptation: ABIDE results},
  author={Li, Xiaoxiao and Gu, Yufeng and Dvornek, Nicha and Staib, Lawrence H and Ventola, Pamela and Duncan, James S},
  journal={Medical image analysis},
  volume={65},
  pages={101765},
  year={2020},
  publisher={Elsevier}
}

@article{10680255,
  title={Brainib: Interpretable brain network-based psychiatric diagnosis with graph information bottleneck},
  author={Zheng, Kaizhong and Yu, Shujian and Li, Baojuan and Jenssen, Robert and Chen, Badong},
  journal={IEEE Transactions on Neural Networks and Learning Systems},
  volume={36},
  number={7},
  pages={13066--13079},
  year={2024},
  publisher={IEEE}
}

@incollection{gkoulalas2015privacy,
  title={Introduction to medical data privacy},
  author={Gkoulalas-Divanis, Aris and Loukides, Grigorios},
  booktitle={Medical data privacy handbook},
  pages={1--14},
  year={2015},
  publisher={Springer}
}

@inproceedings{desai2021continual,
  title={Continual learning with differential privacy},
  author={Desai, Pradnya and Lai, Phung and Phan, NhatHai and Thai, My T},
  booktitle={International Conference on Neural Information Processing},
  pages={334--343},
  year={2021},
  organization={Springer}
}

@inproceedings{rebuffi2017icarl,
  title={icarl: Incremental classifier and representation learning},
  author={Rebuffi, Sylvestre-Alvise and Kolesnikov, Alexander and Sperl, Georg and Lampert, Christoph H},
  booktitle={Proceedings of the IEEE conference on Computer Vision and Pattern Recognition},
  pages={2001--2010},
  year={2017}
}

@article{chaudhry2019tiny,
  title={On tiny episodic memories in continual learning},
  author={Chaudhry, Arslan and Rohrbach, Marcus and Elhoseiny, Mohamed and Ajanthan, Thalaiyasingam and Dokania, Puneet K and Torr, Philip HS and Ranzato, Marc'Aurelio},
  journal={arXiv preprint arXiv:1902.10486},
  year={2019}
}

@inproceedings{kim2024sddgr,
  title={Sddgr: Stable diffusion-based deep generative replay for class incremental object detection},
  author={Kim, Junsu and Cho, Hoseong and Kim, Jihyeon and Tiruneh, Yihalem Yimolal and Baek, Seungryul},
  booktitle={Proceedings of the IEEE/CVF Conference on Computer Vision and Pattern Recognition},
  pages={28772--28781},
  year={2024}
}

@article{buzzega2020dark,
  title={Dark experience for general continual learning: a strong, simple baseline},
  author={Buzzega, Pietro and Boschini, Matteo and Porrello, Angelo and Abati, Davide and Calderara, Simone},
  journal={Advances in neural information processing systems},
  volume={33},
  pages={15920--15930},
  year={2020}
}

@article{li2017learning,
  title={Learning without forgetting},
  author={Li, Zhizhong and Hoiem, Derek},
  journal={IEEE transactions on pattern analysis and machine intelligence},
  volume={40},
  number={12},
  pages={2935--2947},
  year={2017},
  publisher={IEEE}
}

@article{rusu2016progressive,
  title={Progressive neural networks},
  author={Rusu, Andrei A and Rabinowitz, Neil C and Desjardins, Guillaume and Soyer, Hubert and Kirkpatrick, James and Kavukcuoglu, Koray and Pascanu, Razvan and Hadsell, Raia},
  journal={arXiv preprint arXiv:1606.04671},
  year={2016}
}

@inproceedings{
yoon2018lifelong,
title={Lifelong Learning with Dynamically Expandable Networks},
author={Jaehong Yoon and Eunho Yang and Jeongtae Lee and Sung Ju Hwang},
booktitle={International Conference on Learning Representations},
year={2018}
}

@inproceedings{mallya2018packnet,
  title={Packnet: Adding multiple tasks to a single network by iterative pruning},
  author={Mallya, Arun and Lazebnik, Svetlana},
  booktitle={Proceedings of the IEEE conference on Computer Vision and Pattern Recognition},
  pages={7765--7773},
  year={2018}
}

@inproceedings{
chaudhry2019efficient,
title={Efficient Lifelong Learning with A-{GEM}},
author={Arslan Chaudhry and Marc’Aurelio Ranzato and Marcus Rohrbach and Mohamed Elhoseiny},
booktitle={International Conference on Learning Representations},
year={2019}
}

@article{zhang2022hpn,
  title={Hierarchical prototype networks for continual graph representation learning},
  author={Zhang, Xikun and Song, Dongjin and Tao, Dacheng},
  journal={IEEE Transactions on Pattern Analysis and Machine Intelligence},
  volume={45},
  number={4},
  pages={4622--4636},
  year={2022},
  publisher={IEEE}
}

@inproceedings{wang2023sgnngr,
  title={A Generative Adaptive Replay Continual Learning Model for Temporal Knowledge Graph Reasoning},
  author={Zhang, Zhiyu and Chen, Wei and Lin, Youfang and Wan, Huaiyu},
  booktitle={Proceedings of the 63rd Annual Meeting of the Association for Computational Linguistics (Volume 1: Long Papers)},
  pages={10964--10977},
  year={2025}
}

@incollection{mccloskey1989catastrophic,
  title={Catastrophic interference in connectionist networks: The sequential learning problem},
  author={McCloskey, Michael and Cohen, Neal J},
  booktitle={Psychology of learning and motivation},
  volume={24},
  pages={109--165},
  year={1989},
  publisher={Elsevier}
}

@inproceedings{zhou2023ergnn,
  title={Overcoming catastrophic forgetting in graph neural networks with experience replay},
  author={Zhou, Fan and Cao, Chengtai},
  booktitle={Proceedings of the AAAI conference on artificial intelligence},
  volume={35},
  pages={4714--4722},
  year={2021}
}

@article{SHI2024109028,
  title={Continual learning for seizure prediction via memory projection strategy},
  author={Shi, Yufei and Tang, Shishi and Li, Yuxuan and He, Zhipeng and Tang, Shengsheng and Wang, Ruixuan and Zheng, Weishi and Chen, Ziyi and Zhou, Yi},
  journal={Computers in Biology and Medicine},
  volume={181},
  pages={109028},
  year={2024},
  publisher={Elsevier}
}

@article{DUAN2024106338,
  title={Online continual decoding of streaming EEG signal with a balanced and informative memory buffer},
  author={Duan, Tiehang and Wang, Zhenyi and Li, Fang and Doretto, Gianfranco and Adjeroh, Donald A and Yin, Yiyi and Tao, Cui},
  journal={Neural Networks},
  volume={176},
  pages={106338},
  year={2024},
  publisher={Elsevier}
}

@inproceedings{zhou2025brainuicl,
  title={BrainUICL: An unsupervised individual continual learning framework for EEG applications},
  author={Zhou, Yangxuan and Zhao, Sha and Wang, Jiquan and Jiang, Haiteng and Li, Shijian and Li, Tao and Pan, Gang},
  booktitle={The Thirteenth International Conference on Learning Representations},
  year={2025}
}

@article{tzourio2002aal,
  title={Automated anatomical labeling of activations in SPM using a macroscopic anatomical parcellation of the MNI MRI single-subject brain},
  author={Tzourio-Mazoyer, Nathalie and Landeau, Brigitte and Papathanassiou, Dimitri and Crivello, Fabrice and Etard, Octave and Delcroix, Nicolas and Mazoyer, Bernard and Joliot, Marc},
  journal={Neuroimage},
  volume={15},
  number={1},
  pages={273--289},
  year={2002},
  publisher={Elsevier}
}

@inproceedings{10204250,
  title={Data-free knowledge distillation via feature exchange and activation region constraint},
  author={Yu, Shikang and Chen, Jiachen and Han, Hu and Jiang, Shuqiang},
  booktitle={Proceedings of the IEEE/CVF Conference on Computer Vision and Pattern Recognition},
  pages={24266--24275},
  year={2023}
}

@article{song2024llmbasedprivacydataaugmentation,
  title={Llm-based privacy data augmentation guided by knowledge distillation with a distribution tutor for medical text classification},
  author={Song, Yiping and Zhang, Juhua and Tian, Zhiliang and Yang, Yuxin and Huang, Minlie and Li, Dongsheng},
  journal={arXiv preprint arXiv:2402.16515},
  year={2024}
}

@article{zhang2025continualdistillationlearningknowledge,
  title={Continual distillation learning: Knowledge distillation in prompt-based continual learning},
  author={Zhang, Qifan and Guo, Yunhui and Xiang, Yu},
  journal={arXiv preprint arXiv:2407.13911},
  year={2024}
}

@article{10.1145/3702648,
  title={Begin: Extensive benchmark scenarios and an easy-to-use framework for graph continual learning},
  author={Ko, Jihoon and Kang, Shinhwan and Kwon, Taehyung and Moon, Heechan and Shin, Kijung},
  journal={ACM Transactions on Intelligent Systems and Technology},
  volume={16},
  number={1},
  pages={1--22},
  year={2025},
  publisher={ACM New York, NY}
}

@article{Zhang_2025,
  title={Efficient and Robust Continual Graph Learning for Graph Classification in Biology},
  author={Zhang, Ding and Downer, Jane and Chen, Can and Wang, Ren},
  journal={IEEE Transactions on Signal and Information Processing over Networks},
  year={2025},
  publisher={IEEE}
}

@article{Tian2024ContinualLO,
  title={Continual learning on graphs: A survey},
  author={Tian, Zonggui and Zhang, Du and Dai, Hong-Ning},
  journal={arXiv preprint arXiv:2402.06330},
  year={2024}
}

@article{saggar2018towards,
  title={Towards a new approach to reveal dynamical organization of the brain using topological data analysis},
  author={Saggar, Manish and Sporns, Olaf and Gonzalez-Castillo, Javier and Bandettini, Peter A and Carlsson, Gunnar and Glover, Gary and Reiss, Allan L},
  journal={Nature communications},
  volume={9},
  number={1},
  pages={1399},
  year={2018},
  publisher={Nature Publishing Group UK London}
}

@article{logothetis2008we,
  title={What we can do and what we cannot do with fMRI},
  author={Logothetis, Nikos K},
  journal={Nature},
  volume={453},
  number={7197},
  pages={869--878},
  year={2008},
  publisher={Nature Publishing Group UK London}
}

@article{carta2021catastrophicforgettingdeepgraph,
  title={Catastrophic forgetting in deep graph networks: an introductory benchmark for graph classification},
  author={Carta, Antonio and Cossu, Andrea and Errica, Federico and Bacciu, Davide},
  journal={arXiv preprint arXiv:2103.11750},
  year={2021}
}

@article{Kirkpatrick_2017,
  title={Overcoming catastrophic forgetting in neural networks},
  author={Kirkpatrick, James and Pascanu, Razvan and Rabinowitz, Neil and Veness, Joel and Desjardins, Guillaume and Rusu, Andrei A and Milan, Kieran and Quan, John and Ramalho, Tiago and Grabska-Barwinska, Agnieszka and others},
  journal={Proceedings of the national academy of sciences},
  volume={114},
  number={13},
  pages={3521--3526},
  year={2017},
  publisher={National Academy of Sciences}
}

@inproceedings{zenke2017continuallearningsynapticintelligence,
  title={Continual learning through synaptic intelligence},
  author={Zenke, Friedemann and Poole, Ben and Ganguli, Surya},
  booktitle={International conference on machine learning},
  pages={3987--3995},
  year={2017},
  organization={Pmlr}
}

@article{rolnick2019experiencereplaycontinuallearning,
  title={Experience replay for continual learning},
  author={Rolnick, David and Ahuja, Arun and Schwarz, Jonathan and Lillicrap, Timothy and Wayne, Gregory},
  journal={Advances in neural information processing systems},
  volume={32},
  year={2019}
}

@inproceedings{liu2020overcomingcatastrophicforgettinggraph,
  title={Overcoming catastrophic forgetting in graph neural networks},
  author={Liu, Huihui and Yang, Yiding and Wang, Xinchao},
  booktitle={Proceedings of the AAAI conference on artificial intelligence},
  volume={35},
  pages={8653--8661},
  year={2021}
}

@article{hoang2023universalgraphcontinuallearning,
  title={Universal graph continual learning},
  author={Hoang, Thanh Duc and Tung, Do Viet and Nguyen, Duy-Hung and Nguyen, Bao-Sinh and Nguyen, Huy Hoang and Le, Hung},
  journal={arXiv preprint arXiv:2308.13982},
  year={2023}
}

@inproceedings{Zhang_2024,
  title={Topology-aware embedding memory for continual learning on expanding networks},
  author={Zhang, Xikun and Song, Dongjin and Chen, Yixin and Tao, Dacheng},
  booktitle={Proceedings of the 30th ACM SIGKDD Conference on Knowledge Discovery and Data Mining},
  pages={4326--4337},
  year={2024}
}

@article{tan2022graphregularizedmanifoldawareconditionalwasserstein,
  title={Graph-Regularized Manifold-Aware Conditional Wasserstein GAN for Brain Functional Connectivity Generation},
  author={Tan, Yee-Fan and Noman, Fuad and Phan, Rapha{\"e}l C-W and Ombao, Hernando and Ting, Chee-Ming},
  journal={Human Brain Mapping},
  volume={46},
  number={12},
  pages={e70322},
  year={2025},
  publisher={Wiley Online Library}
}

@inproceedings{li2021brainnetgandataaugmentationbrain,
  title={BrainNetGAN: Data augmentation of brain connectivity using generative adversarial network for dementia classification},
  author={Li, Chao and Wei, Yiran and Chen, Xi and Sch{\"o}nlieb, Carola-Bibiane},
  booktitle={MICCAI Workshop on Deep Generative Models},
  pages={103--111},
  year={2021},
  organization={Springer}
}

@article{LIU2021118750,
  title={Graph auto-encoding brain networks with applications to analyzing large-scale brain imaging datasets},
  author={Liu, Meimei and Zhang, Zhengwu and Dunson, David B},
  journal={Neuroimage},
  volume={245},
  pages={118750},
  year={2021},
  publisher={Elsevier}
}

@article{ASD,
  title={The autism brain imaging data exchange: towards a large-scale evaluation of the intrinsic brain architecture in autism},
  author={Di Martino, Adriana and Yan, Chao-Gan and Li, Qingyang and Denio, Erin and Castellanos, Francisco X and Alaerts, Kaat and Anderson, Jeffrey S and Assaf, Michal and Bookheimer, Susan Y and Dapretto, Mirella and others},
  journal={Molecular psychiatry},
  volume={19},
  number={6},
  pages={659--667},
  year={2014},
  publisher={Nature Publishing Group}
}

@article{MDD,
  title={Reduced default mode network functional connectivity in patients with recurrent major depressive disorder},
  author={Yan, Chao-Gan and Chen, Xiao and Li, Le and Castellanos, Francisco Xavier and Bai, Tong-Jian and Bo, Qi-Jing and Cao, Jun and Chen, Guan-Mao and Chen, Ning-Xuan and Chen, Wei and others},
  journal={Proceedings of the National Academy of Sciences},
  volume={116},
  number={18},
  pages={9078--9083},
  year={2019},
  publisher={National Academy of Sciences}
}

@article{BSNIP,
  title={Psychosis biotypes: replication and validation from the B-SNIP consortium},
  author={Clementz, Brett A and Parker, David A and Trotti, Rebekah L and McDowell, Jennifer E and Keedy, Sarah K and Keshavan, Matcheri S and Pearlson, Godfrey D and Gershon, Elliot S and Ivleva, Elena I and Huang, Ling-Yu and others},
  journal={Schizophrenia bulletin},
  volume={48},
  number={1},
  pages={56--68},
  year={2022},
  publisher={Oxford University Press US}
}

@article{kipf2017semisupervisedclassificationgraphconvolutional,
  title={Semi-supervised classification with graph convolutional networks},
  author={Kipf, Thomas N and Welling, Max},
  journal={arXiv preprint arXiv:1609.02907},
  year={2016}
}

@article{velivckovic2017graph,
  title={Graph attention networks},
  author={Veli{\v{c}}kovi{\'c}, Petar and Cucurull, Guillem and Casanova, Arantxa and Romero, Adriana and Lio, Pietro and Bengio, Yoshua},
  journal={arXiv preprint arXiv:1710.10903},
  year={2017}
}

@article{hamilton2018inductiverepresentationlearninglarge,
  title={Inductive representation learning on large graphs},
  author={Hamilton, Will and Ying, Zhitao and Leskovec, Jure},
  journal={Advances in neural information processing systems},
  volume={30},
  year={2017}
}

@article{kipf2016variationalgraphautoencoders,
  title={Variational graph auto-encoders},
  author={Kipf, Thomas N and Welling, Max},
  journal={arXiv preprint arXiv:1611.07308},
  year={2016}
}

@article{dwivedi2021generalizationtransformernetworksgraphs,
  title={A generalization of transformer networks to graphs},
  author={Dwivedi, Vijay Prakash and Bresson, Xavier},
  journal={arXiv preprint arXiv:2012.09699},
  year={2020}
}

@article{yan2016dpabi,
  title={DPABI: data processing \& analysis for (resting-state) brain imaging},
  author={Yan, Chao-Gan and Wang, Xin-Di and Zuo, Xi-Nian and Zang, Yu-Feng},
  journal={Neuroinformatics},
  volume={14},
  number={3},
  pages={339--351},
  year={2016},
  publisher={Springer}
}

@article{ying2021transformersreallyperformbad,
  title={Do transformers really perform badly for graph representation?},
  author={Ying, Chengxuan and Cai, Tianle and Luo, Shengjie and Zheng, Shuxin and Ke, Guolin and He, Di and Shen, Yanming and Liu, Tie-Yan},
  journal={Advances in neural information processing systems},
  volume={34},
  pages={28877--28888},
  year={2021}
}

@article{hoff2002latent,
  title={Latent space approaches to social network analysis},
  author={Hoff, Peter D and Raftery, Adrian E and Handcock, Mark S},
  journal={Journal of the american Statistical association},
  volume={97},
  number={460},
  pages={1090--1098},
  year={2002},
  publisher={Taylor \& Francis}
}


\clearpage
\appendix
\setcounter{page}{1}

\setcounter{figure}{0}
\setcounter{table}{0}
\setcounter{equation}{0}

\begin{center}
    {\fontsize{15}{18}\selectfont\bfseries
    Continual Learning for fMRI-Based Brain Disorder Diagnosis via Functional Connectivity Matrices Generative Replay\par}
    \vspace{0.5em}
    {\fontsize{13}{16}\selectfont Supplementary Material\par}
\end{center}

Our supplementary materials are organized as follows:
\begin{itemize}
  \item Detailed Description of the Cross-Site Continual Learning Pipeline
  \item rs-fMRI Preprocessing
  \item Details in Hierarchical Contextual Thompson Sampling
  \item Extended Evaluation and Ablation Studies of FCM-VAE
  \item Additional Ablation Results
\end{itemize}

\section{Detailed Description of the Cross-Site Continual Learning Pipeline} \label{sec:description}

\begin{figure*}[h]
    \centering
    \includegraphics[width=0.9\textwidth]{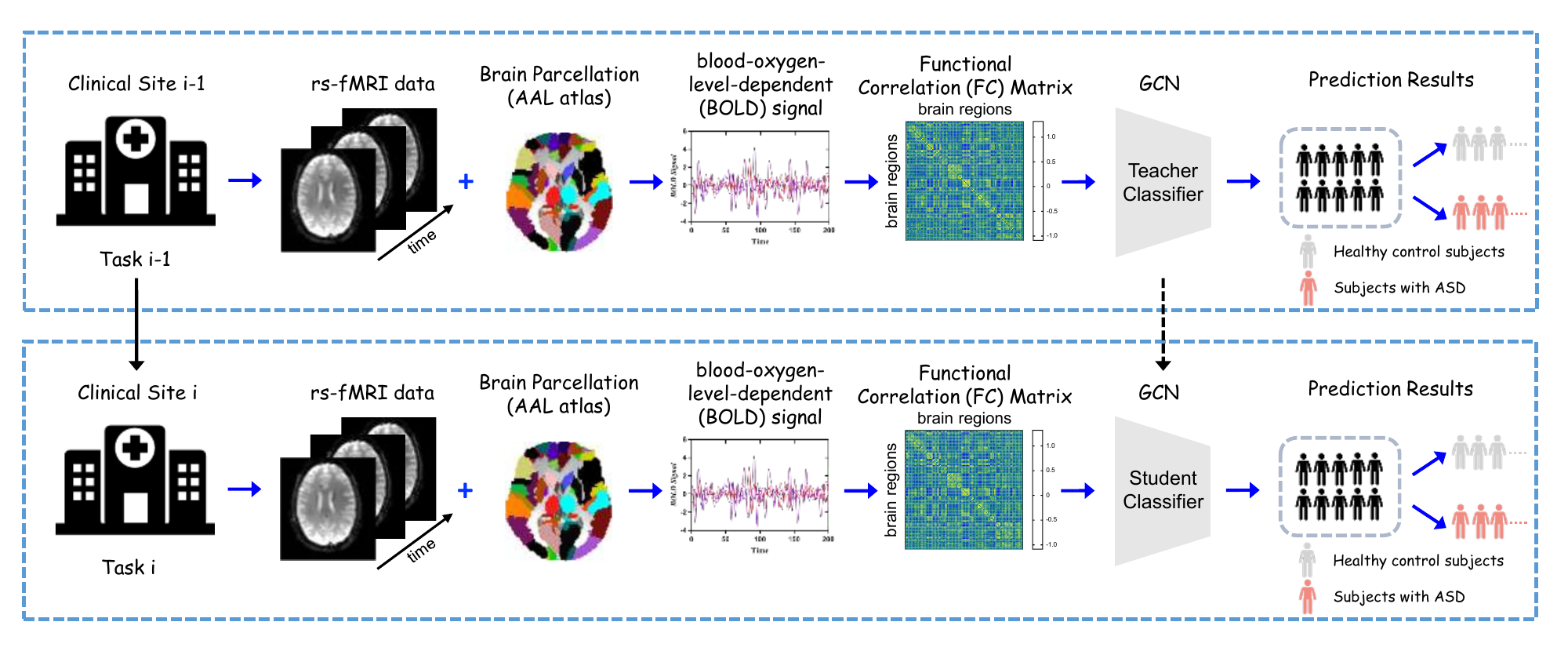}
    \caption{Continual Learning pipeline for transforming raw rs-fMRI data into inputs for the classification task.}
    \label{fig:FC}
\end{figure*}

Figure~\ref{fig:FC} provides an overview of our cross-site continual learning pipeline. 
For each clinical site, we formulate a site-specific classification task that maps rs-fMRI data to diagnostic labels. Starting from the raw rs-fMRI scans acquired at site $i$, we first apply a standardized preprocessing procedure, as described in Section~\ref{sec:preprocessing}. The preprocessed volumes are subsequently parcellated into cortical and subcortical regions using the AAL atlas, yielding a fixed set of $116$ regions of interest (ROIs). For each ROI, we extract its blood-oxygen-level-dependent (BOLD) time series by averaging the voxelwise signals. Functional connectivity (FC) matrices are then computed as the Pearson correlation between every pair of ROI time series, resulting in a subject-specific $116 \times 116$ symmetric FC matrix.

Each FC matrix is interpreted as a weighted brain graph and fed into a graph classifier that outputs a diagnostic prediction. The classification labels distinguish subjects with the target disorder from healthy controls. Across clinical sites, we cast the problem as a continual learning scenario in which each site $i$ defines a task $T_i$ that arrives sequentially.

After completing task $T_{i-1}$, the trained GCN model from site $M_{i-1}$ is frozen and used as a teacher network. When task $T_i$ arrives with data from a new clinical site, we process the rs-fMRI data through the same pipeline and train a student classifier on the new site's FC graphs.

During this stage, the student model is trained jointly on two types of data: (i) the current site's FC matrices and (ii) replayed samples retrieved from the replay buffer. For each training batch, the frozen teacher network produces supervision signals from the same FC inputs, while the student extracts graph embeddings that pass through its prediction head.

\begin{figure}[htbp]
    \centering
    \includegraphics[width=1\linewidth]{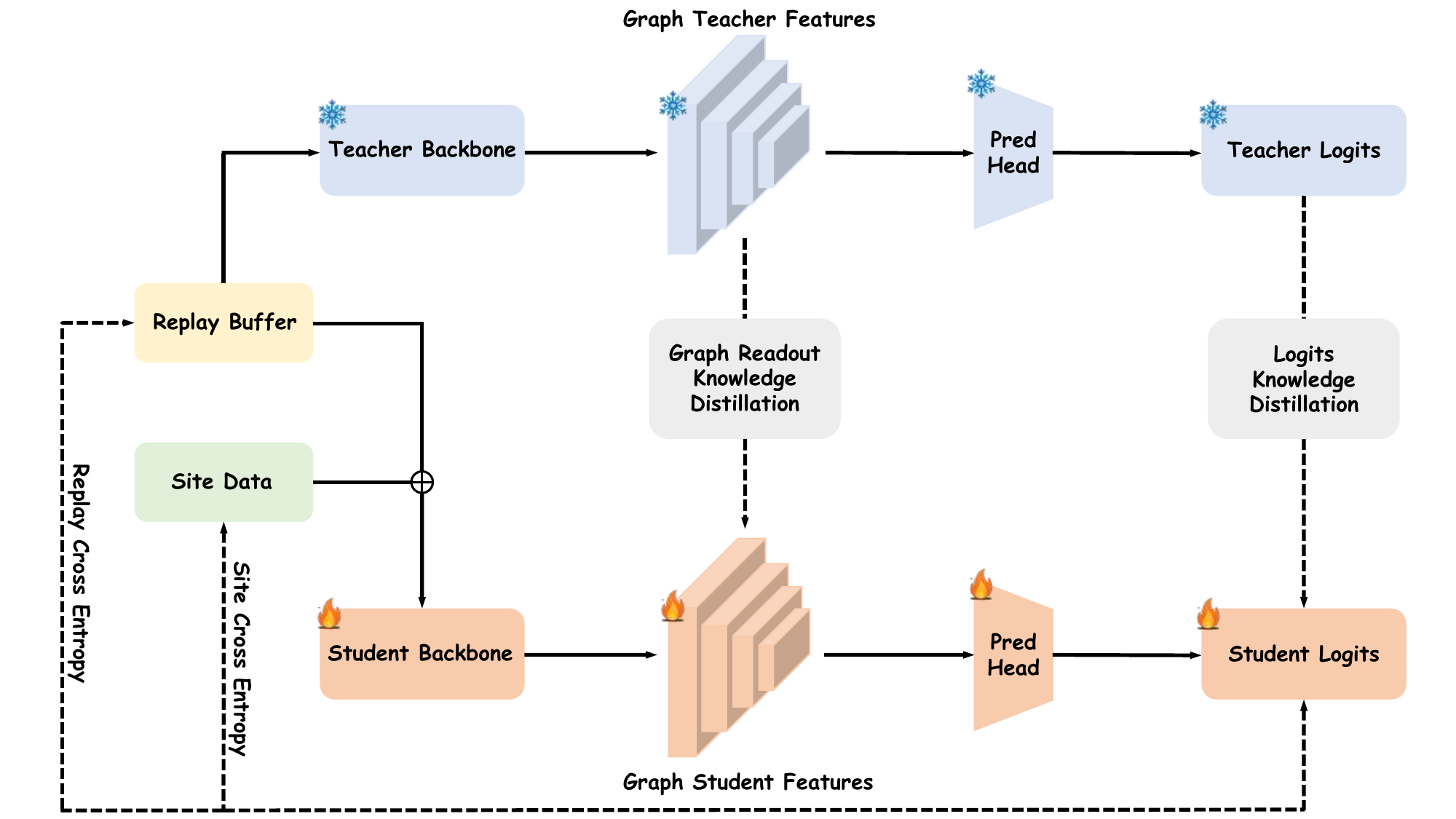}
    \caption{Dual-Level Knowledge Distillation Framework for Continual Graph Learning}
    \label{fig:KD}
\end{figure}

As illustrated in Figure~\ref{fig:KD}, the student model is optimized using three complementary objectives: (i) a graph readout distillation loss that aligns the student's graph-level embeddings with those of the teacher, (ii) a logits distillation loss that enforces agreement between the teacher's and student's predictive distributions, and (iii) a replay loss applied to samples from previous sites. Together, these losses enable the student to retain cross-site knowledge while adapting to newly arriving data, thereby mitigating catastrophic forgetting across tasks.

\section{rs-fMRI Preprocessing}\label{sec:preprocessing}
All rs-fMRI datasets were preprocessed using a standardized pipeline to ensure consistency across clinical sites, following established practices in previous studies. Preprocessing was performed using the FMRIB Software Library (FSL)\footnote{\url{https://fsl.fmrib.ox.ac.uk/fsl/docs/}} and the Statistical Parametric Mapping toolbox (SPM12)\footnote{\url{https://www.fil.ion.ucl.ac.uk/spm/software/spm12/}}. For each subject, the first several volumes were removed to allow for magnetization stabilization. Slice-timing correction was applied to account for inter-slice acquisition delays, followed by rigid-body motion correction to align all volumes to a reference frame and reduce head-motion artifacts. Distortion correction, addressing susceptibility-induced warps using scans with varying acquisition parameters, was conducted with the top-up toolbox in FSL using default settings.   

The preprocessed fMRI volumes were parcellated into anatomically defined regions using the AAL-116 atlas. For each of the 116 ROIs, we extracted the mean BOLD time series by averaging the voxelwise signals within the region. Functional connectivity (FC) matrices were then computed by calculating the Pearson correlation between the time series of every pair of ROIs, resulting in a $116 \times 116$ symmetric FC matrix for each subject.

\section{Details in Hierarchical Contextual Thompson Sampling}\label{sec:HCTS}
While a generative replay buffer ${\mathcal{R}}_i$ enables privacy-preserving rehearsal, using a fixed replay budget is both inefficient and oblivious to distributional shifts across hospitals, ultimately weakening cross-domain generalization. To overcome this, we introduce a Hierarchical Contextual Thompson Sampling mechanism that dynamically allocates replay capacity at two levels: (i) site-level replay quotas adapt to accuracy–forgetting trade-offs; (ii) sample-level quotas enforce diversity and representativeness. Both layers leverage hierarchical contextual features and linear Gaussian posteriors for Thompson sampling, enabling online adaptive optimization. We next detail the sampling strategies at both the site and sample levels.

\subsection{Details of Site-Level Thompson Sampling Allocation}

In cross-hospital continual learning, a fixed replay budget $\mathcal{R}$ restricts each round to a limited number of generated samples. To allocate this budget effectively, we design a site-level adaptive mechanism that jointly accounts for current accuracy and historical forgetting. Concretely, for each site $M_i \in \mathcal{M}$, we define a context vector as:
\[
\boldsymbol{\phi}_i = \big[\mathrm{Acc}_i,\ \mathrm{Forget}_i\big]
\]
where
$
\mathrm{Acc}_i = \tfrac{1}{|\mathcal{D}_i^{\mathrm{test}}|}\!\sum_{(x_i,y_i)\in \mathcal{D}_i^{\mathrm{test}}}\!\mathbbm{1}\{f_\theta(x_i)=y_i\}
$
denotes the current accuracy of $f_\theta$ on site $M_i$, and $\mathbbm{1}$ is the indicator function. $\mathrm{Forget}_i = \max(0,\ \mathrm{Acc}_i^{\mathrm{past}}-\mathrm{Acc}_i^{\mathrm{curr}})$
denotes forgetting at site $M_i$, where $\mathrm{Acc}_i^{\mathrm{past}}$ is the best past accuracy and $\mathrm{Acc}_i^{\mathrm{curr}}$ is the current one.

The expected replay utility is modeled as
\begin{equation}
    r_i \;=\; \boldsymbol{\phi}_i^\top \mathbf{w} + \epsilon_i,
    \qquad \epsilon_i \sim \mathcal{N}(0, \sigma^2),
\end{equation}
where $\mathbf{w}$ is a latent weight vector whose posterior distribution
$p(\mathbf{w}|\mathcal{D})$ encodes uncertainty conditioned on the history $\mathcal{D}$ of previously observed contexts and rewards. At each update round,
Thompson Sampling draws a posterior sample $\tilde{\mathbf{w}} \sim
p(\mathbf{w}|\mathcal{D})$ and computes the predicted utility
$\tilde{r}_i = \boldsymbol{\phi}_i^\top \tilde{\mathbf{w}}$.
Meanwhile, we introduce a closed-loop update mechanism to incorporate performance
feedback. We define the reward change as
\begin{equation}
    r_i^{(t)} \;=\;
    \sigma\!\left(
        \frac{\mathrm{CE}_i^{\mathrm{prev}} - \mathrm{CE}_i^{\mathrm{new}}}{\tau}
    \right),
\end{equation}
where $\sigma(\cdot)$ is the sigmoid function, $\mathrm{CE}_h^{\mathrm{prev}}$ and $\mathrm{CE}_h^{\mathrm{new}}$ are the cross-entropy losses of site $i$ before and after the update, and $\tau$ is a temperature parameter

To maintain a Bayesian estimate of the latent weight vector \(\mathbf{w}\), 
each site keeps sufficient statistics \(A_i\) and \(b_i\), updated at each step as:
\begin{equation}
\begin{aligned}
A_i^{(t+1)} &\leftarrow \gamma A_i^{(t)}
            + (1-\gamma)\,\boldsymbol{\phi}_i^{(t)} {\boldsymbol{\phi}_i^{(t)}}^\top,\\[2mm]
b_i^{(t+1)} &\leftarrow \gamma b_i^{(t)}
            + (1-\gamma)\,\boldsymbol{\phi}_i^{(t)} r_i^{(t)},
\end{aligned}
\end{equation}
where $\gamma$ is a forgetting factor. We initialize $A_i^{(0)} = \lambda I$ and $b_i^{(0)} = 0$ for numerical stability.

The posterior mean and covariance of the Thompson sampling model are approximated as
\begin{equation}
    \mu_i^{(t)} = \big(A_i^{(t)}\big)^{-1} b_i^{(t)}, 
    \qquad
    \Sigma_i^{(t)} = \sigma^2 \big(A_i^{(t)}\big)^{-1}.
\end{equation}

A Thompson sample $\mathbf{w}_i^{(t)} \sim \mathcal{N}\!\left(\mu_i^{(t)},\, \Sigma_i^{(t)}\right)$
yields the predicted utility
\begin{equation}
\hat{r}_i^{(t)} \;=\; {\boldsymbol{\phi}_i^{(t)}}^\top \mathbf{w}_i^{(t)} .
\end{equation}

These predicted utilities are then standardized using \(z\)-score normalization and converted into allocation weights
\begin{equation}
w_i^{(t)} 
= 
\frac{\exp(\hat{r}_i^{(t)})}
     {\sum_{j} \exp(\hat{r}_j^{(t)})} ,
\end{equation}
from which the final replay allocation is obtained via
\begin{equation}
k_i^{(t)} \;=\; K\, w_i^{(t)} .
\end{equation}

After each update, the replay buffer draws \(k_i^{(t)}\) samples from each site's generative memory, the model is updated, and new rewards and contexts are computed, forming a closed-loop adaptive replay strategy.

\subsection{Sample-Level Contextual Thompson Sampling}

Given the site-specific replay quota \(k_i\), we further identify the most beneficial samples within each replay buffer \(\mathcal{R}_i\). To model sample utility, we introduce a sample-level context vector that captures predictive uncertainty and geometric consistency. For a candidate sample \(u\) from site \(M_i\), the context is defined as
\begin{equation}
\label{eq:u-context}
\bm{\psi}_u = \big[\mathrm{margin}_u,\; \mathrm{closeness}_u \big].
\end{equation}

The first component, \(\mathrm{margin}_u\), measures predictive uncertainty. Let \(p=\mathrm{softmax}(h_\theta(u))\) denote the model output probabilities. The margin is computed as
\begin{equation}
\mathrm{margin}_u = \max_{c} p_c - \max_{c' \neq c} p_{c'},
\end{equation}
where a smaller value implies greater ambiguity. Theoretically, low-margin samples generally lie closer to the decision boundary and thus provide higher learning value.

The second component, \(\mathrm{closeness}_u\), quantifies how well \(u\) aligns with its site-specific data manifold. Let 
\[
z_u^{(t)} = \rho\!\left(E_\theta^{(t)}(u)\right)
\]
denote the latent representation at update step \(t\). The closeness score is defined as
\begin{equation}
\begin{aligned}
& \mathrm{closeness}_u
= 1 -  \\
&\tanh \Big(
\sqrt{
\big(z_u^{(t)} - \bm{\mu}_{y_u}^{i,(t)}\big)^\top
\big(\bm{\Sigma}_{y_u}^{i,(t)}\big)^{-1}
\big(z_u^{(t)} - \bm{\mu}_{y_u}^{i,(t)}\big)
}
\Big),
\end{aligned}
\end{equation}
where \(\bm{\mu}_{y_u}^{i,(t)}\) and \(\bm{\Sigma}_{y_u}^{i,(t)}\) are the prototype mean and covariance for class \(y_u\) at site \(M_i\), updated through exponential moving averages over the replay-augmented dataset \(\tilde{\mathcal{D}}_i^{(t)}\). This Mahalanobis-based criterion encourages selecting samples that best preserve site structure while respecting privacy constraints.

To incorporate feedback and maintain Bayesian estimates over sample utilities, each site maintains sufficient statistics \(C_i\) and \(\bm{d}_i\), which capture second-order correlations among sample contexts and their rewards. After selecting a provisional sample subset \(\mathcal{S}_i^{(t)}\), the site-level reward \(r_i^{(t)}\) is evenly assigned to its samples, and the statistics are updated using a forgetting factor \(\gamma\):
\begin{equation}
\begin{aligned}
C_i^{(t)} &\leftarrow 
\gamma C_i^{(t-1)}
+ (1-\gamma)
\sum_{u \in \mathcal{S}_i^{(t)}}
\bm{\psi}_u^{(t)} {\bm{\psi}_u^{(t)}}^\top,\\[4pt]
\bm{d}_i^{(t)} &\leftarrow
\gamma \bm{d}_i^{(t-1)}
+ (1-\gamma)
\sum_{u \in \mathcal{S}_i^{(t)}}
\bm{\psi}_u^{(t)} \frac{r_i^{(t)}}{|\mathcal{S}_i^{(t)}|}.
\end{aligned}
\end{equation}
We initialize \(C_i^{(0)}=\lambda I\) and \(\bm{d}_i^{(0)}=0\) for numerical stability. The posterior mean and covariance are then estimated as
\[
\bm{\mu}_i^{(t)} = (C_i^{(t)})^{-1}\bm{d}_i^{(t)}, 
\quad
\bm{\Sigma}_i^{(t)} = \sigma^2 (C_i^{(t)})^{-1},
\]
where \(\sigma^2\) denotes the reward variance. Thompson sampling is applied by drawing
\[
\mathbf{v}_i^{(t)} \sim \mathcal{N}(\bm{\mu}_i^{(t)}, \bm{\Sigma}_i^{(t)}),
\]
and scoring each candidate \(u\) via
\begin{equation}
s_u^{(t)} = {\mathbf{v}_i^{(t)}}^\top \bm{\psi}_u^{(t)}.
\end{equation}

To select a final replay subset, we combine value-based ranking and diversity promotion. Candidates in \(\tilde{\mathcal{D}}_i^{(t)}\) are first ranked by \(s_u^{(t)}\), and the top \(2k_i\) form a shortlist \(\mathcal{C}_i^{(t)}\). The final \(k_i\) samples are chosen using a greedy farthest-first strategy in the representation space:
\begin{equation}
\label{eq:ffs-final}
\mathcal{S}_i^{(t)} 
= \arg\max_{S \subseteq \mathcal{C}_i^{(t)},\, |S|=k_i}
\min_{\substack{G,G' \in S \\ G \neq G'}}
\left\|
\rho(E_\theta(G)) - \rho(E_\theta(G'))
\right\|_2.
\end{equation}
This ensures that replayed samples are both high-value and distributionally diverse.




\section{Extended Evaluation and Ablation Studies of FCM-VAE}\label{sec:FCM_VAE_more}
\begin{figure*}
    \centering
    \includegraphics[width=1\linewidth]{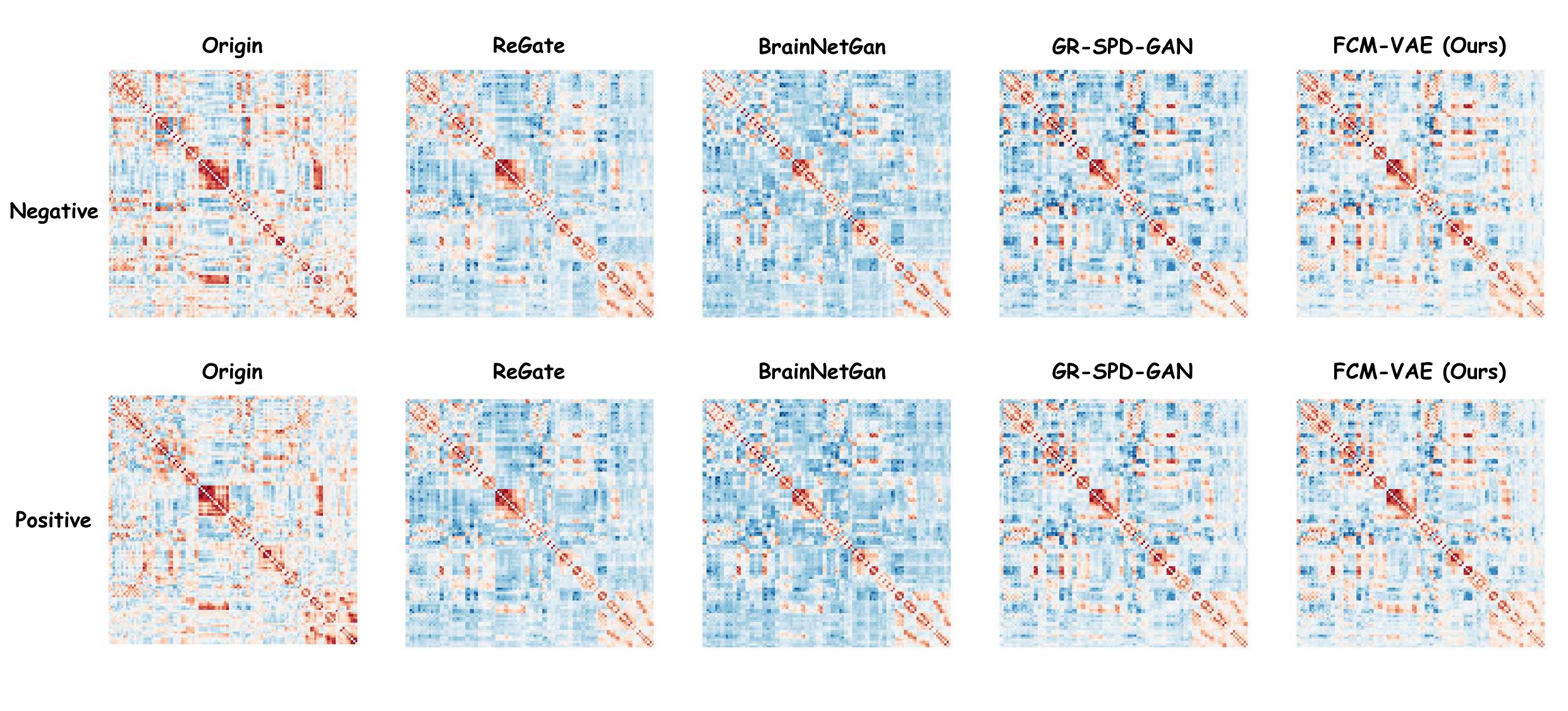}
    \caption{Visualization of functional connectivity (FC) matrices generated by different FC-graph generation methods on the ASD dataset. Columns correspond to different generation method, while rows show negative and positive samples.}
    \label{fig:VAE_results}
\end{figure*}

To better understand the behavior of different FC-graph generators, we first visualize their outputs on the ASD dataset, as shown in Figure~\ref{fig:VAE_results}. For a representative negative and positive subject, we compare the original FC matrix with those produced by ReGate~\citep{LIU2021118750}, BrainNetGAN~\citep{li2021brainnetgandataaugmentationbrain}, GR-SPD-GAN~\citep{tan2022graphregularizedmanifoldawareconditionalwasserstein} and our proposed FCM-VAE. Existing methods roughly recover the coarse modular structure =, but tend to over-smooth fine-grained connectivity patterns or introduce spurious high-magnitude edges, leading to visually distorted or overly homogenized graphs. In contrast, FCM-VAE yields FC matrices that closely follow the original both in global organization and local edge patterns, preserving salient functional modules while suppressing artifacts.

To further validate the robustness of the synthetic FC matrices produced by FCM-VAE, we extend the data augmentation experiment to a broader set of graph classifiers. While the main paper reports results using a GCN backbone, Table~\ref{tab:three-datasets-reordered} evaluates five widely used architectures under different FC generation strategies. Below, we briefly summarize the key modeling characteristics of each classifier:

\begin{table*}[!t]
\centering
\caption{FC classification results with data augmentation using different graph generation methods. 
Each row fixes a classifier and varies the generation model to assess the quality of synthetic FC graphs. 
Columns report accuracy (ACC), precision (PRE), recall (REC), and F1 score on the ASD, BSNIP, and MDD datasets. 
Bold numbers denote the best performance, while underlined numbers indicate the second-best results.}
\label{tab:three-datasets-reordered}
\resizebox{\textwidth}{!}{%
\begin{tabular}{
  l l
  S[table-format=1.3] S[table-format=1.3] S[table-format=1.3] S[table-format=1.3] 
  S[table-format=1.3] S[table-format=1.3] S[table-format=1.3] S[table-format=1.3] 
  S[table-format=1.3] S[table-format=1.3] S[table-format=1.3] S[table-format=1.3] 
}
\toprule
\multirow{2}{*}{\textbf{Classifier}} & \multirow{2}{*}{\textbf{Gen. Method}} &
\multicolumn{4}{c}{\textbf{ASD}} & \multicolumn{4}{c}{\textbf{BSNIP}} & \multicolumn{4}{c}{\textbf{MDD}} \\
\cmidrule(lr){3-6}\cmidrule(lr){7-10}\cmidrule(lr){11-14}
& & \textbf{ACC} & \textbf{PRE} & \textbf{REC} & \textbf{F1}
  & \textbf{ACC} & \textbf{PRE} & \textbf{REC} & \textbf{F1}
  & \textbf{ACC} & \textbf{PRE} & \textbf{REC} & \textbf{F1} \\
\midrule
\multirow{5}{*}{\textbf{GAE}}
& No Augmentation               & \m{\toptwo{0.726}}{\topone{0.753}}{0.669}{0.683} & \m{0.662}{\topone{0.779}}{0.414}{0.421} & \m{\topone{0.767}}{\topone{0.733}}{0.795}{\toptwo{0.731}} \\
& ReGate             & \m{0.711}{0.728}{0.715}{0.689} & \m{0.630}{0.542}{\toptwo{0.421}}{\toptwo{0.434}} & \m{0.722}{0.668}{0.761}{0.688} \\
& BrainNetGan           & \m{0.723}{\toptwo{0.743}}{0.701}{0.696} & \m{0.660}{0.727}{\topone{0.440}}{\topone{0.455}} & \m{0.749}{0.669}{\toptwo{0.816}}{0.726} \\
& GR-SPD-GAN             & \m{0.704}{0.651}{\topone{0.846}}{\toptwo{0.734}} & \m{\toptwo{0.663}}{0.723}{0.377}{\toptwo{0.434}} & \m{0.747}{\toptwo{0.712}}{0.768}{0.710} \\
& \textbf{FCM-VAE (Ours)} & \m{\topone{0.760}}{0.728}{\toptwo{0.831}}{\topone{0.772}} & \m{\topone{0.671}}{\toptwo{0.737}}{0.376}{0.433} & \m{\toptwo{0.765}}{0.648}{\topone{0.885}}{\topone{0.745}} \\
\midrule
\multirow{5}{*}{\textbf{GAT}}
& No Augmentation               & \m{0.713}{0.668}{\topone{0.823}}{0.735} & \m{0.652}{0.698}{0.397}{0.408} & \m{0.683}{0.582}{\topone{0.763}}{\topone{0.653}} \\
& ReGate             & \m{0.683}{0.687}{0.734}{0.674} & \m{0.624}{0.622}{0.315}{0.330} & \m{0.686}{0.589}{\toptwo{0.746}}{\toptwo{0.649}} \\
& BrainNetGan           & \m{0.721}{0.722}{0.716}{0.709} & \m{0.643}{0.684}{\topone{0.444}}{\toptwo{0.442}} & \m{0.683}{0.596}{0.699}{0.635} \\
& GR-SPD-GAN             & \m{\toptwo{0.762}}{\topone{0.768}}{0.735}{\toptwo{0.747}} & \m{0.679}{\topone{0.800}}{\toptwo{0.406}}{0.441} & \m{\toptwo{0.717}}{\toptwo{0.679}}{0.702}{0.600} \\
& \textbf{FCM-VAE (Ours)} & \m{\topone{0.776}}{\toptwo{0.767}}{\toptwo{0.786}}{\topone{0.770}} & \m{\topone{0.690}}{\toptwo{0.765}}{0.403}{\topone{0.493}} & \m{\topone{0.724}}{\topone{0.694}}{0.697}{0.608} \\
\midrule
\multirow{5}{*}{\textbf{GCN}}
& No Augmentation               & \m{0.762}{0.748}{0.772}{0.758} & \m{0.688}{0.733}{0.456}{0.502} & \m{0.756}{\topone{0.733}}{0.699}{0.700} \\
& ReGate             & \m{0.724}{0.688}{\toptwo{0.806}}{0.740} & \m{0.639}{0.618}{0.318}{0.361} & \m{0.724}{0.701}{0.725}{0.676} \\
& BrainNetGan           & \m{0.758}{0.757}{0.752}{0.747} & \m{0.686}{\toptwo{0.747}}{\toptwo{0.458}}{0.488} & \m{0.748}{\toptwo{0.707}}{0.745}{0.712} \\
& GR-SPD-GAN             & \m{\toptwo{0.780}}{\toptwo{0.761}}{\topone{0.812}}{\topone{0.780}} & \m{\toptwo{0.703}}{\topone{0.785}}{0.452}{\toptwo{0.513}} & \m{\toptwo{0.767}}{0.666}{\topone{0.866}}{\topone{0.745}} \\
& \textbf{FCM-VAE (Ours)} & \m{\topone{0.783}}{\topone{0.776}}{0.775}{\toptwo{0.771}} & \m{\topone{0.727}}{0.727}{\topone{0.547}}{\topone{0.597}} & \m{\topone{0.787}}{\topone{0.733}}{\toptwo{0.784}}{\toptwo{0.735}} \\
\midrule
\multirow{5}{*}{\textbf{GraphSage}}
& No Augmentation               & \m{\topone{0.740}}{0.720}{\toptwo{0.784}}{\toptwo{0.736}} & \m{0.656}{0.621}{\toptwo{0.425}}{0.449} & \m{\toptwo{0.751}}{\toptwo{0.712}}{0.707}{\toptwo{0.671}} \\
& ReGate             & \m{0.708}{\topone{0.727}}{0.758}{0.725} & \m{0.623}{0.542}{\topone{0.460}}{\topone{0.467}} & \m{0.725}{0.681}{\toptwo{0.720}}{0.657} \\
& BrainNetGan           & \m{0.695}{0.706}{0.652}{0.658} & \m{0.648}{0.634}{0.365}{0.432} & \m{0.716}{0.678}{0.685}{0.667} \\
& GR-SPD-GAN             & \m{\toptwo{0.729}}{0.716}{0.753}{0.729} & \m{\toptwo{0.658}}{\toptwo{0.644}}{0.365}{\toptwo{0.455}} & \m{0.744}{\topone{0.811}}{0.594}{0.612} \\
& \textbf{FCM-VAE (Ours)} & \m{\topone{0.740}}{\toptwo{0.722}}{\topone{0.795}}{\topone{0.745}} & \m{\topone{0.666}}{\topone{0.702}}{0.357}{0.452} & \m{\topone{0.763}}{0.681}{\topone{0.796}}{\topone{0.730}} \\
\midrule
\multirow{5}{*}{\textbf{GraphTransformer}}
& No Augmentation               & \m{0.694}{0.663}{\topone{0.781}}{\toptwo{0.709}} & \m{0.643}{\topone{0.689}}{0.393}{0.426} & \m{0.692}{0.605}{0.700}{0.636} \\
& Regate             & \m{0.679}{0.653}{\topone{0.781}}{0.700} & \m{0.612}{0.602}{0.260}{0.281} & \m{0.680}{\topone{0.671}}{0.647}{0.571} \\
& BrainNetGan           & \m{0.692}{0.666}{\toptwo{0.766}}{0.705} & \m{0.632}{0.567}{\topone{0.528}}{\topone{0.524}} & \m{0.702}{\toptwo{0.628}}{0.809}{0.678} \\
& GR-SPD-GAN             & \m{\topone{0.725}}{\toptwo{0.708}}{0.756}{\topone{0.720}} & \m{\topone{0.667}}{0.642}{\toptwo{0.480}}{\toptwo{0.507}} & \m{\toptwo{0.719}}{0.609}{\topone{0.859}}{\topone{0.710}} \\
& \textbf{FCM-VAE (Ours)} & \m{\toptwo{0.711}}{\topone{0.716}}{0.684}{0.696} & \m{\toptwo{0.664}}{\toptwo{0.658}}{0.444}{0.459} & \m{\topone{0.723}}{0.627}{\toptwo{0.837}}{\toptwo{0.709}} \\
\bottomrule
\end{tabular}%
}
\end{table*}
\begin{itemize}
    \item \textbf{GAE}~\citep{kipf2016variationalgraphautoencoders}:  
    A graph autoencoder that employs a encoder to map nodes into latent embeddings and an inner-product decoder to reconstruct the graph. 
    It was originally proposed for unsupervised node representation learning and link prediction; in our setting, we use its encoder followed by global pooling as a graph-level classifier.

    \item \textbf{GAT}~\citep{velivckovic2017graph}:  
    A graph attention network where node representations are aggregated using learned attention coefficients, enabling the model to weigh neighbor contributions adaptively based on edge importance.

    \item \textbf{GCN}~\citep{kipf2017semisupervisedclassificationgraphconvolutional}:  
    The standard graph convolutional network that propagates information through normalized message passing. 

    \item \textbf{GraphSAGE}~\citep{hamilton2018inductiverepresentationlearninglarge}:  
     An inductive graph representation learning framework that learns node embeddings by sampling and aggregating information from local neighborhoods using parametric aggregators.
    \item \textbf{GraphTransformer}~\citep{dwivedi2021generalizationtransformernetworksgraphs}:  
    A transformer-based architecture extended to graphs via positional encodings and attention over structural distances.
\end{itemize}

The main architectural and optimization hyperparameters of these classifiers are summarized in Table~\ref{tab:model-zoo-hparams}. 
All models are trained for 100 epochs under identical data splits and optimization settings to ensure a fair comparison.

\begin{table}[htbp]
\centering
\caption{Hyperparameter summary for the graph classifiers used in our experiments.}
\label{tab:model-zoo-hparams}

\footnotesize
\setlength{\tabcolsep}{2.9pt}
\renewcommand{\arraystretch}{1} 
\begin{tabular}{lccccc}
\toprule
\textbf{Model} & \textbf{Layers} & \textbf{Hidden Dim} & \textbf{Dropout} &
\textbf{LR} \\
\midrule

\textbf{GCN} &
4 & [128,128,128,128] & 0.3 &
$1\times10^{-3}$ \\

\textbf{GAT} &
3 & [128,128,128] & 0.4 &
$5\times10^{-4}$ \\

\textbf{GraphSAGE} &
4 & [128,128,128,128] & 0.3 &
$1\times10^{-3}$  \\

\textbf{GraphTransformer} &
3 & [128,128,128] & 0.3 &
$5\times10^{-4}$\\

\textbf{GAE} &
3 & [128,128,128] & 0.2 &
$1\times10^{-3}$\\
\bottomrule
\end{tabular}

\normalsize
\end{table}
\begin{table}[htbp]
\centering
\caption{Grid search ranges for the three hyperparameters 
$\lambda_1$, $\lambda_2$, and $\lambda_3$.}
\begin{tabular}{c c}
\toprule
Coefficient & Search space \\
\midrule
$\lambda_1$ & $\{0.20,\, 0.25,\, 0.30,\, 0.35,\, 0.40\}$ \\
$\lambda_2$ & $\{0.20,\, 0.25,\, 0.30,\, 0.35,\, 0.40\}$ \\
$\lambda_3$ & $\{0.10,\, 0.15,\, 0.20,\, 0.25,\, 0.30\}$ \\
\bottomrule
\end{tabular}
\label{tab:lambda-grid}
\end{table}

For each classifier, we fix the model and vary the FC generation method among five options: no augmentation, ReGate~\citep{LIU2021118750}, BrainNetGAN~\citep{li2021brainnetgandataaugmentationbrain}, GR-SPD-GAN~\citep{tan2022graphregularizedmanifoldawareconditionalwasserstein}, and our FCM-VAE. Performance is reported using accuracy (ACC), precision (PRE), recall (REC), and F1 score on the ASD, BSNIP, and MDD datasets.
\begin{table*}[h]
\centering
\caption{\textbf{Ablation study of FCM-VAE.}
Each row fixes a classifier backbone and varies the ablation setting.
Columns report accuracy (ACC), precision (PRE), recall (REC), and F1 score on the ASD, BRI, and MDD datasets.
Full Framework denotes the complete FCM-VAE model.}
\label{tab:ablation}
\resizebox{\textwidth}{!}{%
\begin{tabular}{
  l l
  S[table-format=1.3] S[table-format=1.3] S[table-format=1.3] S[table-format=1.3] 
  S[table-format=1.3] S[table-format=1.3] S[table-format=1.3] S[table-format=1.3] 
  S[table-format=1.3] S[table-format=1.3] S[table-format=1.3] S[table-format=1.3] 
}
\toprule
\multirow{2}{*}{\textbf{Classifier}} & \multirow{2}{*}{\textbf{Ablation}} &
\multicolumn{4}{c}{\textbf{ASD}} &
\multicolumn{4}{c}{\textbf{BSNIP}} &
\multicolumn{4}{c}{\textbf{MDD}} \\
\cmidrule(lr){3-6}\cmidrule(lr){7-10}\cmidrule(lr){11-14}
& & \textbf{ACC} & \textbf{PRE} & \textbf{REC} & \textbf{F1} 
  & \textbf{ACC} & \textbf{PRE} & \textbf{REC} & \textbf{F1} 
  & \textbf{ACC} & \textbf{PRE} & \textbf{REC} & \textbf{F1} \\
\midrule

\multirow{4}{*}{\textbf{GAE}}
& \textbf{Full Framework} & \m{0.760}{0.728}{0.831}{0.772} & \m{0.671}{0.737}{0.376}{0.433} & \m{0.765}{0.648}{0.885}{0.745}\\
& w/o Node Feature  & \m{0.723}{0.701}{0.769}{0.731} & \m{0.656}{0.732}{0.368}{0.414} & \m{0.710}{0.694}{0.632}{0.582} \\
& w/o LAE           & \m{0.744}{0.714}{0.813}{0.755} & \m{0.664}{0.799}{0.345}{0.390} & \m{0.753}{0.688}{0.776}{0.709} \\
& w/o SGE           & \m{0.734}{0.782}{0.658}{0.706} & \m{0.653}{0.661}{0.379}{0.411} & \m{0.737}{0.618}{0.796}{0.687} \\
\midrule

\multirow{4}{*}{\textbf{GAT}}
& \textbf{Full Framework} & \m{0.776}{0.767}{0.786}{0.770} & \m{0.690}{0.765}{0.403}{0.493} & \m{0.724}{0.694}{0.697}{0.608}\\
& w/o Node Feature  & \m{0.737}{0.761}{0.688}{0.719} & \m{0.658}{0.764}{0.395}{0.420} & \m{0.706}{0.712}{0.580}{0.563} \\
& w/o LAE           & \m{0.743}{0.752}{0.692}{0.720} & \m{0.688}{0.684}{0.459}{0.528} & \m{0.720}{0.688}{0.657}{0.585} \\
& w/o SGE           & \m{0.735}{0.728}{0.741}{0.730} & \m{0.681}{0.739}{0.448}{0.501} & \m{0.697}{0.688}{0.608}{0.563} \\
\midrule

\multirow{4}{*}{\textbf{GCN}}
& \textbf{Full Framework} & \m{0.783}{0.776}{0.775}{0.771} & \m{0.727}{0.727}{0.547}{0.597} & \m{0.787}{0.733}{0.784}{0.735}  \\
& w/o Node Feature  & \m{0.746}{0.762}{0.694}{0.712} & \m{0.693}{0.747}{0.429}{0.489} & \m{0.761}{0.638}{0.887}{0.739} \\
& w/o LAE           & \m{0.762}{0.740}{0.775}{0.754} & \m{0.701}{0.733}{0.439}{0.517} & \m{0.763}{0.705}{0.773}{0.718} \\
& w/o SGE           & \m{0.760}{0.730}{0.786}{0.755} & \m{0.687}{0.635}{0.517}{0.550} & \m{0.745}{0.672}{0.801}{0.720} \\
\midrule

\multirow{4}{*}{\textbf{GraphSAGE}}
& \textbf{Full Framework} & \m{0.740}{0.722}{0.795}{0.745} & \m{0.666}{0.702}{0.357}{0.452} & \m{0.763}{0.681}{0.796}{0.730} \\
& w/o Node Feature  & \m{0.721}{0.727}{0.721}{0.701} & \m{0.644}{0.599}{0.411}{0.474} & \m{0.741}{0.688}{0.713}{0.685} \\
& w/o LAE           & \m{0.738}{0.722}{0.795}{0.745} & \m{0.664}{0.653}{0.394}{0.483} & \m{0.735}{0.694}{0.725}{0.698} \\
& w/o SGE           & \m{0.721}{0.726}{0.708}{0.704} & \m{0.645}{0.704}{0.369}{0.390} & \m{0.742}{0.711}{0.711}{0.694} \\
\midrule

\multirow{4}{*}{\textbf{GraphTransformer}}
& \textbf{Full Framework} & \m{0.711}{0.716}{0.684}{0.696} & \m{0.664}{0.658}{0.444}{0.459} & \m{0.723}{0.627}{0.837}{0.709}\\
& w/o Node Feature  & \m{0.693}{0.685}{0.700}{0.686} & \m{0.651}{0.591}{0.398}{0.426} & \m{0.698}{0.657}{0.713}{0.659} \\
& w/o LAE           & \m{0.708}{0.694}{0.725}{0.702} & \m{0.653}{0.591}{0.484}{0.502} & \m{0.702}{0.662}{0.730}{0.669} \\
& w/o SGE           & \m{0.690}{0.696}{0.672}{0.668} & \m{0.649}{0.602}{0.435}{0.489} & \m{0.698}{0.674}{0.607}{0.619} \\
\bottomrule
\end{tabular}%
}
\end{table*}
Across all classifiers and datasets, FCM-VAE consistently achieves either the best or second-best performance, reflecting its ability to generate FC graphs. Notably, even when paired with simple backbones, FCM-VAE offers substantial improvements over alternative generative models, while GR-SPD-GAN typically ranks second overall. These results reinforce that the advantages of FCM-VAE are not tied to a specific classifier architecture. This demonstrates that incorporating global topological structure and local information into the generative process enhances the quality of the synthesized FC graphs.

We further evaluate the contribution of each component in FCM-VAE by removing them individually. As shown in Table~\ref{tab:ablation}, the absence of any module degrades performance across all datasets, validating their complementary roles. In particular, removing the node feature encoder causes the most significant accuracy drop, emphasizing the importance of incorporating node features with local connectivity patterns and global geometric features. Both the Local Adjacency Encoding (LAE) and Spectral Positional Encoding (SGE) also contribute to preserving local and global structural information, respectively, confirming that their joint design enables more reliable FC graph generation.

\section{Additional Ablation Results}\label{sec:additional_ablation}
\subsection{Component Contributions in FORGE over DER++}
DER++ is a particularly strong and widely adopted continual-learning baseline, making it a rigorous reference point for evaluating FORGE. To clarify what specifically drives FORGE improvements over DER++, we design a controlled ablation that incrementally introduces our key components on top of the DER++ pipeline.

Concretely, starting from DER++ with real sample replay (Real), we replace only the replay source with our generative replay (Gen.) while keeping all other settings fixed. Then, with generative replay held constant, we swap only the sampling strategy (HCTS) or only the distillation objective (Dual-KD). As summarized in Table~\ref{tab:ablation_pipeline}, it shows that HCTS yields the strongest performance, indicating that replay selection is a key driver of the gains, likely because fMRI data quality strongly affects both stability and forgetting. Meanwhile, Dual-KD maintain competitive AAA while achieving lower FOR, suggesting that distillation primarily contributes to reducing forgetting.
\begin{table}[htbp]
\centering
\scriptsize
\setlength{\tabcolsep}{1.2pt}
\renewcommand{\arraystretch}{0.92}
\setlength{\abovecaptionskip}{1pt}
\setlength{\belowcaptionskip}{0pt}
\begin{tabular}{@{}p{0.32\columnwidth} p{0.10\columnwidth} p{0.25\columnwidth} p{0.14\columnwidth} r r@{}}
\hline
\textbf{Method} & \textbf{Replay} & \textbf{Sampling} & \textbf{Distillation} & \textbf{AAA$\uparrow$} & \textbf{FOR$\downarrow$} \\
\hline
\textbf{DER++-Real} & Real & Reservoir Sampling & LKD & 0.669 & 0.187 \\
\textbf{DER++-Gen.} & Gen. & Reservoir Sampling & LKD & 0.671 & 0.183 \\
\textbf{DER++-Gen. + HCTS} & Gen. & HCTS & LKD & 0.685 & 0.141 \\
\textbf{DER++-Gen. + Dual-KD} & Gen. & Reservoir Sampling & Dual-KD & 0.677 & 0.160 \\
\textbf{FORGE} & Gen. & HCTS & Dual-KD & \textbf{0.730} & \textbf{0.116} \\
\hline
\end{tabular}
\caption{Ablation variants by replay, sampling, and distillation.}
\label{tab:ablation_pipeline}
\end{table}

\subsection{Hyperparameter Analysis on regularization weight coefficients}
We study the sensitivity of our continual learning objective by varying the three weights 
$\lambda_1, \lambda_2, \lambda_3$ in:
\begin{equation}
\label{eq:total}
\begin{aligned}
&\mathbb{E}_{(G,y)\sim\mathcal{D}_{t_c}^\mathrm{train}}\!\left[\ell\big(y,f_\theta(G)\big)\right]
+ \lambda_1 \sum_{t=1}^{t_c-1}\mathbb{E}_{(G,y)\sim\mathcal{R}_t}\!\left[\ell\big(y,f_\theta(G)\big)\right]  \\
&\quad + \lambda_2 \sum_{t=1}^{t_c-1} \mathbb{E}_{(G,u)\sim\mathcal{R}_t}
        \left[\left\|h_\theta(G)- u \right\|_2^2\right] \\
& \quad + \lambda_3 \sum_{t=1}^{t_c-1} \mathbb{E}_{(G,r)\sim\mathcal{R}_t}
        \left\|
            \rho\!\big(E_{\theta}(G)\big) - r
        \right\|_2^2,
\end{aligned}
\end{equation}
We perform a grid search over a predefined set of values for these coefficients, with the search ranges informed by parameter choices commonly used in related work~\citep{buzzega2020dark} as summarized in Table~\ref{tab:lambda-grid}. Across all evaluated configurations, the best-performing setting differs slightly
across datasets. Specifically, the optimal coefficients for ASD correspond to 
$\lambda_1 = 0.20$, $\lambda_2 = 0.35$, and $\lambda_3 = 0.15$. 
For the BSNIP dataset, the best configuration is 
$\lambda_1 = 0.25$, $\lambda_2 = 0.30$, and $\lambda_3 = 0.15$, 
while for the MDD dataset the optimal setting becomes 
$\lambda_1 = 0.30$, $\lambda_2 = 0.30$, and $\lambda_3 = 0.20$. 

\begin{figure}[htbp]
    \centering
    \includegraphics[width=0.9\linewidth]{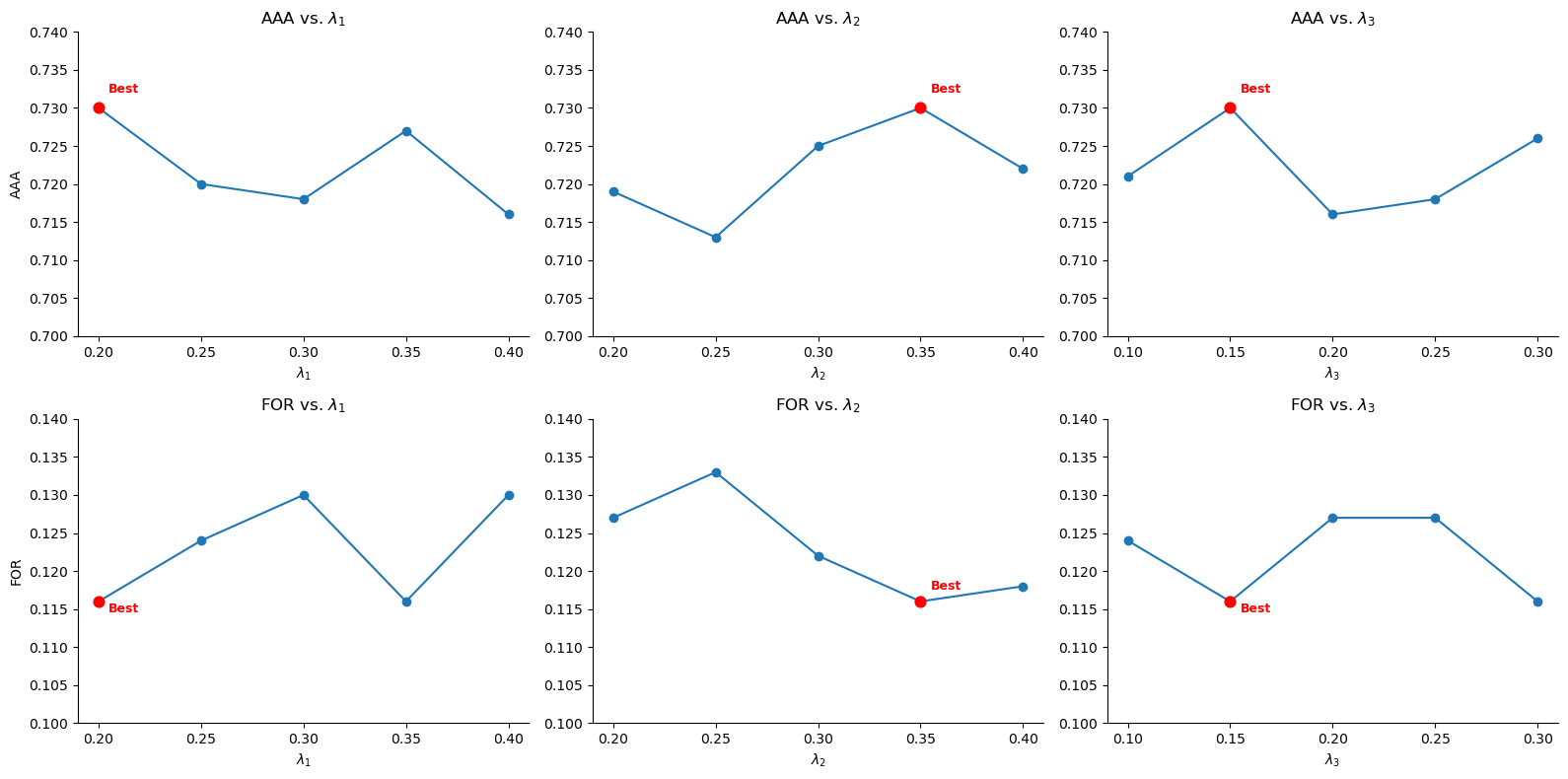}
    \caption{Sensitivity analysis of $\lambda_1$, $\lambda_2$, and $\lambda_3$ on the ASD dataset. For each hyperparameter, we vary its value while keeping the other two coefficients fixed at their default settings.}
    \label{fig:sensitive}
\end{figure}

As illustrated in Figure~\ref{fig:sensitive}, the overall performance remains 
stable as the hyperparameters are varied on the ASD dataset. 
This demonstrates that our framework does not rely on finely tuned loss weights and 
that the model consistently maintains strong accuracy and low forgetting under a wide 
range of $(\lambda_1,\lambda_2,\lambda_3)$ configurations. 
Such behavior highlights the inherent robustness and adaptability of our method,
indicating that strong performance is retained even under large variations in the coefficients.



\end{document}